\documentclass[10pt,final, twocolumn]{IEEEtran}
\usepackage{graphicx,cite,epsfig,amssymb,amsmath}
\usepackage[font=small,labelsep=none, justification=centering]{caption}
\usepackage{algorithm} 
\usepackage{algorithmic}
\usepackage{amsmath}
\usepackage{xcolor}

\begin{document}
\markboth{IEEE Journal on Selected Areas in Communications, Vol. XX, No. Y,
Month 2015} {Zi etc.: Energy Efficiency Optimization of 5G Radio Frequency Chain Systems \ldots}
\title{\mbox{}\vspace{0.40cm}\\
\textsc{Energy Efficiency Optimization of 5G Radio Frequency Chain Systems} \vspace{0.2cm}}

\author{\normalsize
%

Ran Zi$^1$, Xiaohu Ge$^1$, John Thompson$^2$, Cheng-Xiang Wang$^3$, Haichao Wang$^1$, Tao Han$^1$\\
\vspace{0.70cm} \small{
$^1$Department of Electronics and Information Engineering\\
Huazhong University of Science and Technology, Wuhan 430074, Hubei, P. R.
China.\\}
Email: \{xhge,hantao\}@mail.hust.edu.cn, 1192408534@qq.com\\
\vspace{0.1cm}
$^2$Institute for Digital Communications, \\
University of Edinburgh, Edinburgh, EH9 3JL, UK.\\
Email: john.thompson@ed.ac.uk\\
\vspace{0.1cm}
$^3$Joint Research Institute for Signal and Image Processing, \\
School of Engineering \& Physical Sciences, \\
Heriot-Watt University, Edinburgh, EH14 4AS, UK.\\
Email: cheng-xiang.wang@hw.ac.uk\\

\vspace{-0.5cm}
\thanks{\scriptsize{ Submitted to IEEE Journal on Selected Areas in Communications Special Issue on Energy-Efficient Techniques for 5G Wireless Communication Systems.}}
\thanks{\scriptsize{Correspondence author: Dr. Xiaohu Ge, Tel: +86 27 8755 7942, Fax: +86 27 8755 7943, Email: xhge@mail.hust.edu.cn }}
\thanks{\scriptsize{The authors would like to acknowledge the support from the International Science and Technology Cooperation Program of China  under grants 2015DFG12580 and 2014DFA11640, the National Natural Science Foundation of China (NSFC) under the grants 61471180, NFSC Major International Joint Research Project under the grant 61210002, the Fundamental Research Funds for the Central Universities under the grant 2015XJGH011. This research is partially supported by the EU FP7-PEOPLE-IRSES, project acronym S2EuNet (grant no. 247083), project acronym WiNDOW (grant no. 318992) and project acronym CROWN (grant no. 610524), the EU H2020 5G Wireless project (Grant no. 641985), EU FP7 QUICK project (Grant no. PIRSES-GA-2013-612652), National 863 project in 5G by Ministry of Science and Technology in China (Grant no. 2014AA01A701), National international Scientific and Technological Cooperation Base of Green Communications and Networks (No. 2015B01008) and Hubei International Scientific and Technological Cooperation Base of Green Broadband Wireless Communications.}}
}

\date{\today}
\renewcommand{\baselinestretch}{1.2}
\thispagestyle{empty} \maketitle \thispagestyle{empty}
\newpage
\setcounter{page}{1}\begin{abstract}

With the massive multi-input multi-output (MIMO) antennas technology adopted for the fifth generation (5G) wireless communication systems, a large number of radio frequency (RF) chains have to be employed for RF circuits. However, a large number of RF chains not only increase the cost of RF circuits but also consume additional energy in 5G wireless communication systems. In this paper we investigate energy and cost efficiency optimization solutions for 5G wireless communication systems with a large number of antennas and RF chains. An energy efficiency optimization problem is formulated for 5G wireless communication systems using massive MIMO antennas and millimeter wave technology. Considering the non-concave feature of the objective function, a suboptimal iterative algorithm, i.e., the energy efficient hybrid precoding (EEHP) algorithm is developed for maximizing the energy efficiency of 5G wireless communication systems. To reduce the cost of RF circuits, the energy efficient hybrid precoding with the minimum number of RF chains (EEHP-MRFC) algorithm is also proposed. Moreover, the critical number of antennas searching (CNAS) and user equipment number optimization (UENO) algorithms are further developed to optimize the energy efficiency of 5G wireless communication systems by the number of transmit antennas and UEs. Compared with the maximum energy efficiency of conventional zero-forcing (ZF) precoding algorithm, numerical results indicate that the maximum energy efficiency of the proposed EEHP and EEHP-MRFC algorithms are improved by 220\% and 171\%, respectively.

\end{abstract}
\IEEEpeerreviewmaketitle \vspace{-1cm}

\section{Introduction}

The massive multi-input multi-output (MIMO) antennas and the millimeter wave communication technologies have been widely known as two key technologies for the fifth generation (5G) wireless communication systems \cite{Thompson14, Boccardi14,Swindlehurst14,cxwang14,Chen15,Chen15_2}. Compared with conventional MIMO antenna technology, massive MIMO can improve more than 10 times spectrum efficiency in wireless communication systems \cite{Marzetta10}. Moreover, the beamforming gain based on the massive MIMO antenna technology helps to overcome the path loss fading in millimeter wave channels. For MIMO communication systems with traditional radio frequency (RF) chains and baseband processing, one antenna corresponds to one RF chain \cite{Vu07, Zhang05}. In this case, a large number of RF chains has to be employed for massive MIMO communication systems. These RF chains not only consume a large amount of energy in wireless transmission systems but also increase the cost of wireless communication systems \cite{Emil15}. Therefore, it is an important problem to find energy efficient solutions for 5G wireless communication systems with a large number of antennas and RF chains.


To improve the performance of multiple antenna transmission systems, hybrid precoding technology combining digital baseband precoding with analog RF precoding was investigated in \cite{Zhang05,El14, Bogale14, Alkhateeb14, Ahmed14, Liu14, Liang14, Tadilo14}. Based on the joint design of RF chains and baseband processing, a soft antenna subset selection scheme was proposed for multiple antenna channels \cite{Zhang05}. When a single data stream is transmitted, the soft antenna subset selection scheme was able to achieve the same signal-to-noise ratio (SNR) gain as a full-complexity scheme involving all antennas in MIMO wireless communication systems. By formulating the problem of millimeter wave precoder design as a sparsity-constrained signal recovery problem, algorithms were developed to approximate optimal unconstrained precoders and combiners in millimeter wave communication systems with large antenna arrays \cite{El14}. To maximize the sum rate of MIMO communication systems, a hybrid beamforming approach was designed indirectly by considering a weighted sum mean square error minimization problem incorporating the solution of digital beamforming systems \cite{Bogale14}. Based on a low-complexity channel estimation algorithm, a hybrid precoding algorithm was proposed to achieve a near-optimal performance relative to the unconstrained digital solutions in the single user millimeter wave communication system \cite{Alkhateeb14}. To reduce the feedback overhead in millimeter wave MIMO communication systems, a low complexity hybrid analog/digital precoding scheme was developed for the downlink of the multi-user communication systems \cite{Ahmed14}. To maximize the minimum average data rate of users subject to a limited RF chain constraint and a phase-only constraint, a two stage precoding scheme was proposed to exploit the large spatial degree of freedom gain in massive MIMO systems with reduced channel state information (CSI) signaling overhead \cite{Liu14}. Based on phase-only constraints in the RF domain and a low-dimensional baseband zero-forcing (ZF) precoding, a low complexity hybrid precoding scheme was presented to approach the performance of the traditional baseband ZF precoding scheme \cite{Liang14}. Compared with the digital beamforming scheme, the hybrid beamforming scheme was shown to achieve the same performance with the minimum RF chains and phase shifters in multi-user massive MIMO communication systems \cite{Tadilo14}. In the above studies, hybrid precoding schemes have rarely been investigated to optimize the energy efficiency of massive MIMO communication systems. While in cellular communication systems, energy efficiency has been considered as a critical performance metric \cite{Xiang13}.

To improve the energy efficiency of MIMO communication systems, some digital precoding schemes have been studied in \cite{Belmega11,Jiang13,Xu13}. Based on static and fast-fading MIMO channels, an energy efficient precoding scheme was investigated when the terminals are equipped with multiple antennas \cite{Belmega11}. Jointly considering the transmit power, power allocation among date streams and beamforming matrices, a power control and beamforming algorithm was developed for MIMO interference channels to maximize the energy efficiency of communication systems \cite{Jiang13}. Transforming the energy efficiency of MIMO broadcast channel into a concave fractional program, an optimization approach with transmit covariance optimization and active transmit antenna selection was proposed to improve the energy efficiency of MIMO systems over broadcast channels \cite{Xu13}. {With the massive MIMO concept emerging as a key technology in 5G communication systems, the energy efficiency of massive MIMO has been studied in several papers \cite{Emil15,Ngo13,Ng12,Yang13,Ha13,Mohammed14}. Based on a new power consumption model for multi-user massive MIMO, closed-form expressions involving the number of antennas, number of active users and gross rate were derived for maximizing the energy efficiency of massive MIMO systems with ZF processing \cite{Emil15}. When linear precoding schemes are adopted at the BS, it is proved that massive MIMO can improve the energy efficiency by three orders of magnitude \cite{Ngo13}. Considering the transmit power and the circuit power in massive MIMO systems, a power consumption model has been proposed to help optimize the energy efficiency of multi-cell mobile communication systems by selecting the optimal number of active antennas \cite{Ha13}. When the sum spectrum efficiency was fixed, the impact of transceiver power consumption on the energy efficiency of ZF detector was investigated for the uplinks of massive MIMO systems \cite{Mohammed14}. Considering the power cost by RF generation, baseband computing, and the circuits associated with each antenna, simulation results in \cite{Yang13} illustrated that massive MIMO macro cells outperforms LTE macro cells in both spectrum and energy efficiency. To maximize the energy efficiency of the massive MIMO OFDMA systems, an energy efficient iterative algorithm was proposed by optimizing the power allocation, data rate, antenna number, and subcarrier allocation in \cite{Ng12}.}


However, in all the aforementioned studies, only the transmission rate and the hardware complexity of hybrid precoding systems were analyzed for MIMO or massive MIMO communication systems. Moreover, energy efficient solutions for the RF chains and the baseband processing of 5G wireless communication systems is surprisingly rare in the open literature. On the other hand, the energy and cost increase through using a large number of RF chains is an inevitable problem for 5G wireless communication systems. Motivated by the above gaps, in this paper we propose energy and cost efficient optimization solutions for 5G wireless communication systems with a large number of antennas and RF chains. The contributions and novelties of this paper are summarized as follows.

\begin{enumerate}
\item The BS energy efficiency including the energy consumption of RF chains and baseband processing is formulated as an optimization function for 5G wireless communication systems adopting massive MIMO antennas and millimeter wave technologies.
\item Considering the non-concave feature of the optimization objective function, a suboptimal solution is proposed to maximize the BS energy efficiency using the energy efficient hybrid precoding (EEHP) algorithm.
\item To reduce the cost of RF circuits, the energy efficient hybrid precoding with the minimum number of RF chains (EEHP-MRFC) algorithm is developed to tradeoff the energy and cost efficiency for 5G wireless communication systems.
\item To utilize the user scheduling and resource management schemes, the critical number of antennas searching (CNAS) and user equipment number optimization (UENO) algorithms are developed to maximize the energy efficiency of 5G wireless communication systems.
\end{enumerate}

The remainder of this paper is outlined as follows. Section II describes the system model of 5G wirelss communication systems and the energy efficiency optimization problem is formulated. In Section III, an iteration algorithm, i.e., the EEHP algorithm is developed to maximize the energy efficiency of 5G wireless communication systems. To save the cost of RF circuits, the EEHP-MRFC algorithm is developed to tradeoff the energy and cost efficiency in Section IV. Moreover, the CNAS and UENO algorithms are developed to optimize the energy efficiency of 5G wireless communication systems considering the number of transmit antennas and user equipments (UEs). Detailed numerical simulations are presented in Section V. Finally, conclusions are drawn in Section VI.

\section{System model}

Considering the impact of massive MIMO antennas on the RF chains and the baseband processing, the energy efficiency of 5G wireless communication systems has to be rethought. In the following, we describe the system configuration of 5G wireless communication systems, including massive MIMO antennas, RF chains and the baseband processing. Moreover, the energy efficiency optimization problem of 5G wireless communication systems is formulated.

\subsection{System Configuration}

\begin{figure}
\vspace{0.1in}
\centerline{\includegraphics[width=9cm,draft=false]{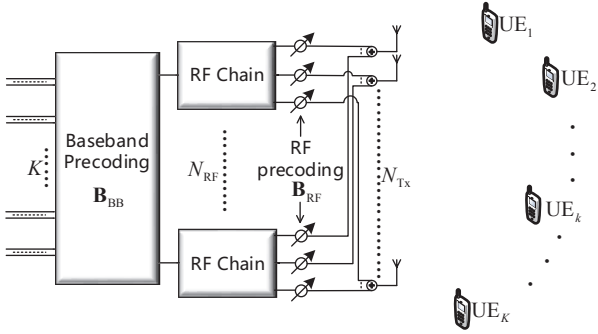}}
\caption{\small \ \ system model.}
\label{fig1}
\end{figure}

Without loss of generality, a single cell scenario is illustrated in Fig.~\ref{fig1}, where a BS and $K$ UEs are located in the 5G wireless communication system. The BS is assumed to be equipped with ${N_{{\text{Tx}}}}$ antennas. And based on the massive MIMO configuration, we assume ${N_{{\text{Tx}}}} \geqslant 100$ in our system \cite{Marzetta10,Ngo13}. There are $K$ active UEs, each with a single antenna, that are associated with the BS. Moreover, the transmission system of BS is equipped with ${N_{{\text{RF}}}}$ RF chains. One baseband data stream is assumed to be associated with one UE in 5G wireless communication systems. In this paper, our studies focuses on the downlinks of 5G wireless communication systems.

The received signal at the $k{\text{th}}$ UE is expressed as
\begin{equation}
{y_k} = {\bf{h}}_k^H{{\bf{B}}_{{\rm{RF}}}}{{\bf{B}}_{{\rm{BB}}}}{\bf{x}} + {w_k},
\label{eq1}
\tag{1}
\end{equation}
where ${\mathbf{x}} = {\left[ {{x_1},...,{x_k},...,{x_K}} \right]^H}$ is the signal vector transmitted from the BS to $K$ UEs, where the ${x_k}, k = 1,...,K$, are assumed to be independently and identically distributed ({\em i.i.d.}) Gaussian random variables with zero mean and variance of 1; ${{\mathbf{B}}_{{\text{BB}}}} \in {\mathbb{C}^{{N_{{\text{RF}}}} \times K}}$ is the baseband precoding matrix, where the $k{\text{th}}$ column of ${{\mathbf{B}}_{{\text{BB}}}}$ is denoted as ${{\mathbf{b}}_{{\text{BB}},k}}$ which is the baseband precoding vector associated with the $k{\text{th}}$ UE; ${{\mathbf{B}}_{{\text{RF}}}} \in {\mathbb{C}^{{N_{{\text{Tx}}}} \times {N_{{\text{RF}}}}}}$ is the RF precoding matrix which is performed by ${N_{{\text{RF}}}}$ RF chains; ${w_k}$ is the noise received by the $k{\text{th}}$ UE. Moreover, all noises received by UEs are denoted as {\em i.i.d.} Gaussian random variables with zero mean and variance of 1. The vector ${\mathbf{h}}_k^H$ is the downlink channel vector between the BS and the $k{\text{th}}$ UE. The downlink channel matrix between the BS and $K$ UEs is denoted as ${{\mathbf{H}}^H} = {\left[ {{{\mathbf{h}}_1},...,{{\mathbf{h}}_k},...,{{\mathbf{h}}_K}} \right]^H}$. The power consumed to transmit signals for the $k{\text{th}}$ UE is expressed as
\begin{equation}
{P_k} = {\left\| {{{\mathbf{B}}_{{\text{RF}}}}{{\mathbf{b}}_{{\text{BB}},k}}{x_k}} \right\|^2} = {\left\| {{{\mathbf{B}}_{{\text{RF}}}}{{\mathbf{b}}_{{\text{BB}},k}}} \right\|^2}.
\label{eq2}
\tag{2}
\end{equation}

{Note that the consumed power for transmitting signals in (2) is expended for a given bandwidth $W$, which is set to be 20MHz in this paper \cite{Emil15}.} The millimeter wave communication technology is adopted for 5G wireless communication systems. Considering the propagation characteristic of millimeter waves in wireless communications, a geometry-based stochastic modeling (GBSM) is used to express the millimeter wave channel as follows \cite{El14,Bogale14,Xu02,Raghavan11}
\begin{equation}
{{\mathbf{h}}_k} = \sqrt {\frac{{{N_{{\text{Tx}}}}{\beta _k}}}{{{N_{{\text{ray}}}}}}} \sum\limits_{i = 1}^{{N_{{\text{ray}}}}} {{\rho _{ki}}{\mathbf{u}}\left( {{\psi _i},{\vartheta _i}} \right)},
\label{eq3}
\tag{3}
\end{equation}
where ${N_{{\text{ray}}}}$ is the number of the multipath between the BS and $K$ UEs. ${\beta _k} = {\zeta  \mathord{\left/
 {\vphantom {\zeta  {l_k^\gamma }}} \right.
 \kern-\nulldelimiterspace} {l_k^\gamma }}$ is the large scale fading coefficient over the wireless link between the BS and the $k{\text{th}}$ UE. $\zeta $ is the lognormal random variable with the zero mean and the variance of 9.2 dB. ${l_k}$ is the distance between the BS and the $k{\text{th}}$ UE. $\gamma $ is the path loss exponent. {${\rho _{ki}}$ is the complex gain of the $i{\text{th}}$ multipath over the $k{\text{th}}$ UE link, which denotes the small-scale fading in wireless channels and is governed by a complex Gaussian distribution. Moreover, ${\rho _{ki}}$ is i.i.d. for different values of $k (k = 1,...,K)$ and $i (i = 1,...,{{N_{{\text{ray}}}}})$.} ${\psi _i}$ and ${\vartheta _i}$ are the azimuth and the elevation angle of the $i{\text{th}}$ multipath at the BS antenna array, respectively. ${\mathbf{u}}\left( {{\psi _i},{\vartheta _i}} \right)$ is the response vector of BS antenna array with the azimuth ${\psi _i}$ and the elevation angle ${\vartheta _i}$. Without loss of generality, the BS antenna array is assumed as the uniform planar antenna array in this paper. Therefore, the response vector of BS antenna array with the azimuth ${\psi _i}$ and the elevation angle ${\vartheta _i}$ is expressed as \cite{Balanis12}
\begin{equation}
\begin{gathered}
  {\mathbf{u}}\left( {{\psi _i},{\vartheta _i}} \right) = \frac{1}{{\sqrt {{N_{{\text{Tx}}}}} }}\left[ {1,.} \right...,{e^{j\frac{{2\pi }}{\lambda }d\left( {m\sin \left( {{\psi _i}} \right)sin\left( {{\vartheta _i}} \right) + ncos\left( {{\vartheta _i}} \right)} \right)}}, \hfill \\
  \;\;\;\;\;\;\;\;\;...,{\left. {{e^{j\left( {{N_{{\text{Tx}}}} - 1} \right)\frac{{2\pi }}{\lambda }d\left( {\left( {M - 1} \right)\sin \left( {{\psi _i}} \right)sin\left( {{\vartheta _i}} \right) + \left( {N - 1} \right)cos\left( {{\vartheta _i}} \right)} \right)}}} \right]^T} \hfill \\
\end{gathered},
\label{eq4}
\tag{4}
\end{equation}
where $d$ is the distance between adjacent antennas, $\lambda $ is the carrier wave length, $M$ and $N$ are the row and column number of the BS antenna array, respectively. $m$ is denoted as the $m{\text{th}}$ antenna in the row of the BS antenna array, $1 \leqslant m < M$; $n$ is denoted as the $n{\text{th}}$ antenna in the column of the BS antenna array, $1 \leqslant n < N$.

\subsection{Problem Formulation}

Based on the system model in Fig.~\ref{fig1}, the link spectrum efficiency of the $k{\text{th}}$ UE is expressed as

\begin{equation}
{R_k} = {\log _2}\left( {1 + \frac{{{\mathbf{h}}_k^H{{\mathbf{B}}_{{\text{RF}}}}{{\mathbf{b}}_{{\text{BB}},k}}{\mathbf{b}}_{{\text{BB}},k}^H{\mathbf{B}}_{{\text{RF}}}^H{{\mathbf{h}}_k}}}{{\sum\limits_{i = 1,i \ne k}^K {{\mathbf{h}}_k^H{{\mathbf{B}}_{{\text{RF}}}}{{\mathbf{b}}_{{\text{BB}},i}}{\mathbf{b}}_{{\text{BB}},i}^H{\mathbf{B}}_{{\text{RF}}}^H{{\mathbf{h}}_k}}  + \sigma _n^2}}} \right).
\label{eq5}
\tag{5}
\end{equation}

{Note that from (5) we get the instantaneous spectrum efficiency. And in this paper, we assume that the BS transmitter has perfect CSI, i.e. channel vectors ${{{\mathbf{h}}_k}}, k=1,...,K$ are known at the BS. This assumption is widely adopted for the investigation of precoding problems in massive MIMO systems and millimeter wave transmission systems \cite{El14,Bogale14,Ngo13,Ha13}. In practical wireless communication systems, the CSI can be obtained through uplink channel estimation then applied to downlink precoding based on the channel reciprocity in the time division duplex (TDD) mode \cite{Marzetta10,Bogale14CISS}. Moreover, the millimeter wave multipath channel estimation utilizing compressed channel sensing was investigated in \cite{Alkhateeb14} and \cite{Bajwa10}.} Furthermore, considering all the UEs, the sum spectrum efficiency is expressed by
\begin{equation}
{R_{{\text{sum}}}} = \sum\limits_{k = 1}^K {{R_k}}.
\label{eq6}
\tag{6}
\end{equation}

The total BS power is expressed as
\begin{equation}
{P_{{\text{total}}}} = \frac{1}{\alpha }\sum\limits_{k = 1}^K {{{\left\| {{{\mathbf{B}}_{{\text{RF}}}}{{\mathbf{b}}_{{\text{BB}},k}}} \right\|}^2}}  + {N_{{\text{RF}}}}{P_{{\text{RF}}}} + {P_{\text{C}}},
\label{eq7}
\tag{7}
\end{equation}
where $\alpha $ is the efficiency of the power amplifier, {and the term $\sum\limits_{k = 1}^K {{{\left\| {{{\mathbf{B}}_{{\text{RF}}}}{{\mathbf{b}}_{{\text{BB}},k}}} \right\|}^2}} $ is the power consumed to transmit signals for $K$ UEs over the given bandwidth $W$,} ${P_{{\text{RF}}}}$ is the power consumed at every RF chain which is comprised by converters, mixers, filters, phase shifters, etc. Considering the number of antennas is fixed in a wireless communication system, the number of phase shifters is also fixed since each phase shifter is associated with one antenna. ${P_{\text{C}}}$ is the power consumed for site-cooling, baseband processing and synchronization in the BS. To simplify the derivation, ${P_{\text{C}}}$ is fixed as a constant.

In this paper, we focus on how to maximize the BS energy efficiency (bits per Joule) by optimizing the baseband precoding matrix ${{\mathbf{B}}_{{\text{BB}}}}$, the RF precoding matrix ${{\mathbf{B}}_{{\text{RF}}}}$ and the number of RF chains ${N_{{\text{RF}}}}$. This optimization problem is formed by
\begin{equation}
\begin{gathered}
  \left( {N_{{\text{RF}}}^{{\text{opt}}},{\mathbf{B}}_{{\text{RF}}}^{{\text{opt}}},{\mathbf{B}}_{{\text{BB}}}^{{\text{opt}}}} \right) = \mathop {\arg \;\max }\limits_{{N_{{\text{RF}}}},{{\mathbf{B}}_{{\text{RF}}}},\;\;{{\mathbf{B}}_{{\text{BB}}}}} \;\eta  = \frac{{W{R_{{\text{sum}}}}}}{{{P_{{\text{total}}}}}} \hfill \\
  \;\;\;\;\;\;\;\;\;\;\;\;\;\;\;\;\;\;\;\;\;\;\;\;\;\;\;{\text{s}}{\text{.t}}{\text{.}}\;\;\;{\left| {{{\left[ {{{\mathbf{B}}_{{\text{RF}}}}} \right]}_{i,j}}} \right|^2} = \frac{1}{{{N_{{\text{Tx}}}}}} \hfill \\
  \;\;\;\;\;\;\;\;\;\;\;\;\;\;\;\;\;\;\;\;\;\;\;\;\;\;\;\;\;\;\;\;\;{R_k} \geqslant {\Gamma _k},\;k = 1,...,K \hfill \\
  \;\;\;\;\;\;\;\;\;\;\;\;\;\;\;\;\;\;\;\;\;\;\;\;\;\;\;\;\;\;\;\;\;\sum\limits_{k = 1}^K {{{\left\| {{{\mathbf{B}}_{{\text{RF}}}}{{\mathbf{b}}_{{\text{BB}},k}}} \right\|}^2}}  \leqslant {P_{{\text{max}}}} \hfill \\
\end{gathered},
\label{eq8}
\tag{8}
\end{equation}
where $W$ is the transmission bandwidth, $\Gamma _k$ is the minimum spectrum efficiency required by ${\text{U}}{{\text{E}}_k}$, and ${P_{{\text{max}}}}$ is the maximum transmit power required for wireless downlinks. In general, the RF precoding is performed by phase shifters, which can change the signal phase but can not change the signal amplitude. Therefore, the amplitude of RF precoding matrix is fixed as a constant, which is added as a constraint for (\ref{eq8}), i.e., ${\left| {{{\left[ {{{\mathbf{B}}_{{\text{RF}}}}} \right]}_{i,j}}} \right|^2} = \frac{1}{{{N_{{\text{Tx}}}}}}$. {Furthermore, there exists a minimum data rate required by each UE in practical applications. The minimum data rate is obtained by multiplying the corresponding minimum spectrum efficiency by the bandwidth. Since the bandwidth is fixed in this paper, an equivalent minimum spectrum efficiency constraint is used to satisfy the minimum data rate requirement, which is expressed as ${R_k} \geqslant {\Gamma _k},\;k = 1,...,K$ in (8).} And $\sum\limits_{k = 1}^K {{{\left\| {{{\mathbf{B}}_{{\text{RF}}}}{{\mathbf{b}}_{{\text{BB}},k}}} \right\|}^2}}  \leqslant {P_{{\text{max}}}} $ is the maximum transmit power constraint.

\section{Energy Efficient Hybrid Precoding Design}

To maximize the energy efficiency in (\ref{eq8}), an EEHP algorithm is developed to jointly optimize the baseband precoding matrix, the RF precoding matrix and the number of RF chains in the following.

\begin{algorithm*}[!t]
\caption{\textbf{Energy Efficient Hybrid Precoding-A (EEHP-A) algorithm}.}
\label{alg1}
\begin{algorithmic}
\STATE \textbf{Begin:} \begin{enumerate}
                        \item Assuming ${{\mathbf{B}}^{\left( 0 \right)}}$ to be the initial digital precoding matrix, ${\mathbf{\Omega }}_k^{\left( 0 \right)}$ and ${\mathbf{\Xi }}_k^{\left( 0 \right)}$ are calculated by (10) for the UE ${\text{U}}{{\text{E}}_k},{\text{ }}k = 1,...,K$;
    \item At the $n{\text{th}}$, $n = 1,2,...$, iteration step, the iterative step length $\mu _k^{\left( n \right)}$ is searched within $\left[ {0,1} \right]$ for all $K$ UEs
    \begin{equation}
   	\begin{gathered}
  \mu _k^{\left( n \right)} = \arg \mathop {\max }\limits_{\mu _k^{\left( n \right)} \in [0,1]} \eta \left\{ {\left[ {{{\mathbf{I}}_{{N_{{\text{Tx}}}}}} + \mu _k^{\left( n \right)}\left( {{{\left[ {{\mathbf{\Xi }}_k^{\left( {n - 1} \right)}} \right]}^{ - 1}}{\mathbf{\Omega }}_k^{\left( {n - 1} \right)} - {{\mathbf{I}}_{{N_{{\text{Tx}}}}}}} \right)} \right]{\mathbf{b}}_k^{\left( {n - 1} \right)}} \right\} \hfill \\
  \;\;\;\;\;\;\;\;{\text{s}}{\text{.t}}{\text{.}}\;\;{{\bar R}_k}\left( {\left[ {{{\mathbf{I}}_{{N_{{\text{Tx}}}}}} + \mu _k^{\left( n \right)}\left( {{{\left[ {{\mathbf{\Xi }}_k^{\left( {n - 1} \right)}} \right]}^{ - 1}}{\mathbf{\Omega }}_k^{\left( {n - 1} \right)} - {{\mathbf{I}}_{{N_{{\text{Tx}}}}}}} \right)} \right]{\mathbf{b}}_k^{\left( {n - 1} \right)}} \right) \geqslant {\Gamma _k}\;\; \hfill \\
  \;\;\;\;\;\;\;\;\;\;\;\;\;\sum\limits_{k = 1}^K {{{\left\| {\left[ {{{\mathbf{I}}_{{N_{{\text{Tx}}}}}} + \mu _k^{\left( n \right)}\left( {{{\left[ {{\mathbf{\Xi }}_k^{\left( {n - 1} \right)}} \right]}^{ - 1}}{\mathbf{\Omega }}_k^{\left( {n - 1} \right)} - {{\mathbf{I}}_{{N_{{\text{Tx}}}}}}} \right)} \right]{\mathbf{b}}_k^{\left( {n - 1} \right)}} \right\|}^2}}  \leqslant {P_{{\text{max}}}} \hfill \\
  \end{gathered} ;
	\label{eq12}
	\tag{12}
	\end{equation}

    \item Based on $\mu _k^{\left( n \right)}$, the digital precoding matrix ${\mathbf{b}}_k^{\left( n \right)}$ is calculated at the $n{\text{th}}$ $n = 1,2,...$  iteration step
        \begin{equation}
   	{\mathbf{b}}_k^{\left( n \right)} = \left[ {{{\mathbf{I}}_{{N_{{\text{Tx}}}}}} + \mu _k^{\left( n \right)}\left( {{{\left[ {{\mathbf{\Xi }}_k^{\left( {n - 1} \right)}} \right]}^{ - 1}}{\mathbf{\Omega }}_k^{\left( {n - 1} \right)} - {{\mathbf{I}}_{{N_{{\text{Tx}}}}}}} \right)} \right]{\mathbf{b}}_k^{\left( {n - 1} \right)};
	\label{eq13}
	\tag{13}
	\end{equation}

    \item Based on ${\mathbf{b}}_k^{\left( n \right)}$ and (10), ${\mathbf{\Omega }}_k^{\left( n \right)}$ and ${\mathbf{\Xi }}_k^{\left( n \right)}$ are updated;

    \item Return to step 2 and keep iterating till ${\mathbf{b}}_k^{\left( n \right)}$ converges.
                       \end{enumerate}
\STATE \textbf{end Begin}
\end{algorithmic}
\end{algorithm*}

\subsection{Upper Bound of Energy Efficiency}

Based on definitions of the baseband precoding matrix and the RF precoding matrix, the size of ${{\mathbf{B}}_{{\text{RF}}}} \in {\mathbb{C}^{{N_{{\text{Tx}}}} \times {N_{{\text{RF}}}}}}$ and ${{\mathbf{B}}_{{\text{BB}}}} \in {\mathbb{C}^{{N_{{\text{RF}}}} \times K}}$ is related with the number of RF chains ${N_{{\text{RF}}}}$. To simplify the derivation, we first fix the number of RF chains. Based on the optimization objective function in the Section II, (\ref{eq8}) is a non-concave function with regard to ${{\mathbf{B}}_{{\text{RF}}}}$ and ${{\mathbf{B}}_{{\text{BB}}}}$. In general, there does not exist an analytical solution for such non-concave functions. To tackle this problem, ${\mathbf{B}} = {{\mathbf{B}}_{{\text{RF}}}}{{\mathbf{B}}_{{\text{BB}}}}$ is configured as a digital precoding matrix ${\mathbf{B}} \in {\mathbb{C}^{{N_{{\text{Tx}}}} \times K}}$, ${{\mathbf{B}}} = {\left[ {{{\mathbf{b}}_1},...,{{\mathbf{b}}_k},...,{{\mathbf{b}}_K}} \right]}$ whose size does not depend on the number of RF chains ${N_{{\text{RF}}}}$. Furthermore, the amplitude of ${\mathbf{B}}$ is free from the constraint ${\left| {{{\left[ {{{\mathbf{B}}_{{\text{RF}}}}} \right]}_{i,j}}} \right|^2} = \frac{1}{{{N_{{\text{Tx}}}}}}$. Substituting the digital precoding matrix ${\mathbf{B}}$ into (\ref{eq8}) and neglecting the constraint ${\left| {{{\left[ {{{\mathbf{B}}_{{\text{RF}}}}} \right]}_{i,j}}} \right|^2} = \frac{1}{{{N_{{\text{Tx}}}}}}$, the optimization problem in the Section II is transformed as
\begin{equation}
\begin{gathered}
  {{\mathbf{B}}^{{\text{opt}}}} = \mathop {\arg \;\max }\limits_{\mathbf{B}} \;\bar \eta  = \frac{{W\sum\limits_{k = 1}^K {{{\bar R}_k}} }}{{\frac{1}{\alpha }\sum\limits_{k = 1}^K {{{\left\| {{{\mathbf{b}}_k}} \right\|}^2}}  + {N_{{\text{RF}}}}{P_{{\text{RF}}}} + {P_{\text{C}}}}} \hfill \\
  \;\;\;\;\;\;\;\;\;\;\;\;\;\;\;\;\;\;\;\;\;\;\;\;\;\;\;{\text{s}}{\text{.t}}{\text{.}}\;\;\;{{\bar R}_k} \geqslant {\Gamma _k},\;k = 1,...,K \hfill \\
  \;\;\;\;\;\;\;\;\;\;\;\;\;\;\;\;\;\;\;\;\;\;\;\;\;\;\;\;\;\;\;\;\;\sum\limits_{k = 1}^K {{{\left\| {{{\mathbf{b}}_k}} \right\|}^2}}  \leqslant {P_{{\text{max}}}} \hfill \\
\end{gathered},
\label{eq9a}
\tag{9a}
\end{equation}
where
\begin{equation}
{\bar R_k} = {\log _2}\left( {1 + \frac{{{\mathbf{h}}_k^H{{\mathbf{b}}_k}{\mathbf{b}}_k^H{{\mathbf{h}}_k}}}{{\sum\limits_{i = 1,i \ne k}^K {{\mathbf{h}}_k^H{{\mathbf{b}}_i}{\mathbf{b}}_i^H{{\mathbf{h}}_k}}  + \sigma _n^2}}} \right).
\label{eq9b}
\tag{9b}
\end{equation}

Based on (9), the energy efficient is maximized by optimizing the digital precoding matrix ${\mathbf{B}}$, where ${\mathbf{B}}$ is only constrained by the maximum transmit power ${P_{{\text{max}}}}$ and the minimum spectrum efficiency $\Gamma _k$. Assume that the maximum energy efficiency in (\ref{eq8}) is denoted as ${\eta ^{\max }}$. Similarly, assume that the maximum energy efficiency in (9) is denoted as ${\bar \eta ^{\max }}$. Compared (\ref{eq8}) with (9), two optimization problems have the same objective function. But (9) has less constraints than that of (\ref{eq8}). Therefore, the solution of (9), i.e., the maximum energy efficiency of (9) is larger than or equal to the solution of (\ref{eq8}), i.e., the maximum energy efficiency of (8). Therefore, the maximum energy efficiency ${\eta ^{\max }}$ is upper bounded by the maximum energy efficiency ${\bar \eta ^{\max }}$, i.e., ${\eta ^{\max }} \leqslant {\bar \eta ^{\max }}$.

\subsection{Energy Efficiency Local Optimization}

Considering (9) is a non-concave function, it is difficult to find a global optimization solution for this optimization problem. A local optimization solution ${\bar \eta ^{{\text{opt}}}}$ with the optimal digital precoding matrix ${{\mathbf{B}}^{{\text{opt}}}} = \left[ {{\mathbf{b}}_1^{{\text{opt}}}, \cdots ,{\mathbf{b}}_k^{{\text{opt}}}, \cdots ,{\mathbf{b}}_K^{{\text{opt}}}} \right]$ is first derived for the energy efficiency optimization in (9).

Denoting the energy efficiency and spectrum efficiency in (9) as functions of ${{\mathbf{b}}_k}$, i.e. $\bar \eta \left( {{{\mathbf{b}}_k}} \right)$ and ${\bar R_k}\left( {{{\mathbf{b}}_k}} \right)$. The gradient of $\bar \eta \left( {{{\mathbf{b}}_k}} \right)$ with respect to ${{\mathbf{b}}_k}$ is derived by
\begin{equation}
\frac{{\partial \bar \eta \left( {{{\mathbf{b}}_k}} \right)}}{{\partial {{\mathbf{b}}_k}}} = \frac{2}{{{{\bar P}^2}}}\left[ {{{\mathbf{\Omega }}_k} - {{\mathbf{\Xi }}_k}} \right]{{\mathbf{b}}_k},
\label{eq10a}
\tag{10a}
\end{equation}
with
\begin{equation}
{{\mathbf{\Omega }}_k} = \frac{{\bar P{{\mathbf{h}}_k}{\mathbf{h}}_k^H}}{{\sum\limits_{j = 1}^K {{\mathbf{h}}_k^H{{\mathbf{b}}_j}{\mathbf{b}}_j^H{{\mathbf{h}}_k}}  + \sigma _n^2}},
\label{eq10b}
\tag{10b}
\end{equation}

\begin{equation}
{{\mathbf{\Xi }}_{k}}=\frac{\sum\limits_{i=1}^{K}{{{{\bar{R}}}_{i}}}}{\alpha \ln 2}{{\mathbf{I}}_{{{N}_{\text{Tx}}}}}+\bar{P}\sum\limits_{i=1,i\ne k}^{K}{\left( \frac{\mathbf{h}_{i}^{H}{{\mathbf{b}}_{i}}\mathbf{b}_{i}^{H}{{\mathbf{h}}_{i}}}{{{\left( {{\delta }_{i}} \right)}^{2}}+{{\delta }_{i}}\mathbf{h}_{i}^{H}{{\mathbf{b}}_{i}}\mathbf{b}_{i}^{H}{{\mathbf{h}}_{i}}}\centerdot {{\mathbf{h}}_{i}}\mathbf{h}_{i}^{H} \right)},
\label{eq10c}
\tag{10c}
\end{equation}

\begin{equation}
\bar P = \frac{1}{W}\left( {\frac{1}{\alpha }\sum\limits_{k = 1}^K {{{\left\| {{{\mathbf{b}}_k}} \right\|}^2}}  + {N_{{\text{RF}}}}{P_{{\text{RF}}}} + {P_{\text{C}}}} \right),
\label{eq10d}
\tag{10d}
\end{equation}

\begin{equation}
{\delta _i} = \sum\limits_{j = 1,j \ne i}^K {{\mathbf{h}}_i^H{{\mathbf{b}}_j}{\mathbf{b}}_j^H{{\mathbf{h}}_i}}  + \sigma _n^2,
\label{eq10e}
\tag{10e}
\end{equation}

where ${\delta _i}$ is the sum of the received interference power and the noise power for the UE ${\text{U}}{{\text{E}}_i},{\text{ }}i = 1,...,K$. ${\bar R_i}$ is obtained from (9b) by replacing $k$ with $i$.

When the zero-gradient condition $\frac{{\partial \bar \eta \left( {{{\mathbf{b}}_k}} \right)}}{{\partial {{\mathbf{b}}_k}}} = 0$ is applied, the local optimization solution for the UE ${\text{U}}{{\text{E}}_k},{\text{ }}k = 1,...,K$ is derived as
\begin{equation}
{{\mathbf{\Xi }}_k}{{\mathbf{b}}_k} = {{\mathbf{\Omega }}_k}{{\mathbf{b}}_k}.
\label{eq11}
\tag{11}
\end{equation}

To obtain the optimal digital precoding matrix ${{\mathbf{B}}^{{\text{opt}}}}$ of the local optimization solution, an iterative algorithm is developed and called the energy efficiency hybrid precoding-A (EEHP-A) algorithm on the top of the previous page.

Based on the EEHP-A algorithm, the ${\mathbf{b}}_k^{{\text{opt}}}$ is obtained by the converged ${\mathbf{b}}_k^{\left( n \right)}$. After obtaining ${\mathbf{b}}_k^{{\text{opt}}}{\text{, }}k = 1, \cdots ,K$ for all $K$ UEs, the local optimization solution ${\bar \eta ^{{\text{opt}}}}$ is achieved. To ensure the convergence of EEHP-A algorithm, the corresponding proof is given as follows.

\textbf{Proof }: Assuming ${\mathbf{X}} = {\left[ {{\mathbf{\Xi }}_k^{\left( {n - 1} \right)}} \right]^{ - 1}}{\mathbf{\Omega }}_k^{\left( {n - 1} \right)}{\mathbf{b}}_k^{\left( {n - 1} \right)}$, (\ref{eq12}) and (\ref{eq13}) are rewritten as
 \begin{equation}
\begin{gathered}
  \mu _k^{\left( n \right)} = \arg \mathop {\max }\limits_{\mu _k^{\left( n \right)} \in [0,1]} \eta \left\{ {\mu _k^{\left( n \right)}{\mathbf{X}} + \left( {1 - \mu _k^{\left( n \right)}} \right){\mathbf{b}}_k^{\left( {n - 1} \right)}} \right\} \hfill \\
  \;\;\;\;\;\;\;\;{\text{s}}{\text{.t}}{\text{.}}\;\;{{\bar R}_k}\left( {{\mathbf{b}}_k^{\left( n \right)}} \right) \geqslant {\Gamma _k} \hfill \\
  \;\;\;\;\;\;\;\;\;\;\;\;\;\sum\limits_{k = 1}^K {{{\left\| {{\mathbf{b}}_k^{\left( n \right)}} \right\|}^2}}  \leqslant {P_{{\text{max}}}} \hfill \\
\end{gathered},
	\label{eq14}
	\tag{14}
	\end{equation}

 \begin{equation}
   	\begin{gathered}
  {\mathbf{b}}_k^{\left( n \right)} = \mu _k^{\left( n \right)}{\left[ {{\mathbf{\Xi }}_k^{\left( {n - 1} \right)}} \right]^{ - 1}}{\mathbf{\Omega }}_k^{\left( {n - 1} \right)}{\mathbf{b}}_k^{\left( {n - 1} \right)} + \left( {1 - \mu _k^{\left( n \right)}} \right){\mathbf{b}}_k^{\left( {n - 1} \right)} \hfill \\
  \;\;\;\;\;\;\;\; = \mu _k^{\left( n \right)}{\mathbf{X}} + \left( {1 - \mu _k^{\left( n \right)}} \right){\mathbf{b}}_k^{\left( {n - 1} \right)} \hfill \\
\end{gathered} .
	\label{eq15}
	\tag{15}
	\end{equation}
:

For the UE ${\text{U}}{{\text{E}}_k},{\text{ }}k = 1,...,K$, ${\mathbf{\Xi }}_k^{\left( {n - 1} \right)}$ based on (\ref{eq10c}) is a Hermitian symmetric positive matrix. Hence, ${\mathbf{\Xi }}_k^{\left( {n - 1} \right)}$ can be denoted as ${\mathbf{\Xi }}_k^{\left( {n - 1} \right)} = {\mathbf{Z}}{{\mathbf{Z}}^H}$, where ${\mathbf{Z}}$ is a symmetric positive definite matrix. Furthermore, the following result is derived as (16):

\begin{equation}
   	\begin{gathered}
  {\left[ {\frac{{\partial \bar \eta \left( {{\mathbf{b}}_k^{\left( {n - 1} \right)}} \right)}}{{\partial {\mathbf{b}}_k^{\left( {n - 1} \right)}}}} \right]^H}\left( {{\mathbf{X}} - {\mathbf{b}}_k^{\left( {n - 1} \right)}} \right) \hfill \\
   = \frac{2}{{{{\bar P}^2}}}{\left[ {{\mathbf{b}}_k^{\left( {n - 1} \right)}} \right]^H}\left( {{\mathbf{\Omega }}_k^{\left( {n - 1} \right)} - {\mathbf{\Xi }}_k^{\left( {n - 1} \right)}} \right) \times  \hfill \\
  {\kern 1pt} {\kern 1pt} {\kern 1pt} {\kern 1pt} {\kern 1pt} {\kern 1pt} {\kern 1pt} {\kern 1pt} {\kern 1pt} {\kern 1pt} {\kern 1pt} {\kern 1pt} \left( {{{\left[ {{\mathbf{\Xi }}_k^{\left( {n - 1} \right)}} \right]}^{ - 1}}{\mathbf{\Omega }}_k^{\left( {n - 1} \right)} - 1} \right){\mathbf{b}}_k^{\left( {n - 1} \right)} \hfill \\
   = \frac{2}{{{{\bar P}^2}}}{\left[ {{\mathbf{b}}_k^{\left( {n - 1} \right)}} \right]^H}\left( {{\mathbf{\Omega }}_k^{\left( {n - 1} \right)}{{\left[ {{\mathbf{\Xi }}_k^{\left( {n - 1} \right)}} \right]}^{ - 1}}{\mathbf{\Omega }}_k^{\left( {n - 1} \right)} - } \right. \hfill \\
  {\kern 1pt} {\kern 1pt} {\kern 1pt} {\kern 1pt} {\kern 1pt} {\kern 1pt} {\kern 1pt} {\kern 1pt} {\kern 1pt} {\kern 1pt} {\kern 1pt} {\kern 1pt} {\kern 1pt} {\kern 1pt} \left. {2{\mathbf{\Omega }}_k^{\left( {n - 1} \right)} + {\mathbf{\Xi }}_k^{\left( {n - 1} \right)}} \right){\mathbf{b}}_k^{\left( {n - 1} \right)} \hfill \\
   = \frac{2}{{{{\bar P}^2}}}{\left[ {{\mathbf{b}}_k^{\left( {n - 1} \right)}} \right]^H}{\left( {{{\mathbf{Z}}^{ - 1}}{\mathbf{\Omega }}_k^{\left( {n - 1} \right)} - {\mathbf{\Omega }}_k^{\left( {n - 1} \right)}} \right)^H} \times  \hfill \\
  \;\;\;\left( {{{\mathbf{Z}}^{ - 1}}{\mathbf{\Omega }}_k^{\left( {n - 1} \right)} - {\mathbf{\Omega }}_k^{\left( {n - 1} \right)}} \right){\mathbf{b}}_k^{\left( {n - 1} \right)}\underset{\raise0.3em\hbox{$\smash{\scriptscriptstyle-}$}}{ \succ } 0 \hfill \\
\end{gathered} .
	\label{eq16}
	\tag{16}
	\end{equation}

Based on (\ref{eq16}) and the proposition in \cite{Jiang11}, $\bar \eta \left( {{\mathbf{b}}_k^{\left( 0 \right)}} \right) \leqslant \bar \eta \left( {{\mathbf{b}}_k^{\left( 1 \right)}} \right) \leqslant  \cdots  \leqslant \bar \eta \left( {{\mathbf{b}}_k^{\left( n \right)}} \right)$ are non-decreasing sequences for the UE ${\text{U}}{{\text{E}}_k},{\text{ }}k = 1,...,K$. Moreover, $\bar \eta \left( {{{\mathbf{b}}_k}} \right)$ is upper-bounded according to the proposition in \cite{Jiang13}. It is easily known that the upper-bounded non-decreasing sequence approaches to a convergence value. Therefore, $\bar \eta \left( {{{\mathbf{b}}_k}} \right)$ in EEHP-A is proved to converge.

\subsection{Hybrid Precoding Matrices Optimization}

Based on (9) and the EEHP-A algorithm, the local optimization solution ${\bar \eta ^{{\text{opt}}}}$ is obtained. When the hybrid precoding matrices ${{\mathbf{B}}_{{\text{RF}}}}{{\mathbf{B}}_{{\text{BB}}}}$ approach the optimal digital precoding matrix ${{\mathbf{B}}^{{\text{opt}}}}$ with the constraint ${\left| {{{\left[ {{{\mathbf{B}}_{{\text{RF}}}}} \right]}_{i,j}}} \right|^2} = \frac{1}{{{N_{{\text{Tx}}}}}}$, the energy efficiency $\eta $ will approaches the local optimization solution ${\bar \eta ^{{\text{opt}}}}$. Therefore, the optimal hybrid precoding matrices ${\mathbf{B}}_{{\text{RF}}}^{{\text{opt}}}$ and ${\mathbf{B}}_{{\text{BB}}}^{{\text{opt}}}$ can be solved by minimizing the Euclidean distance between ${{\mathbf{B}}_{{\text{RF}}}}{{\mathbf{B}}_{{\text{BB}}}}$ and ${{\mathbf{B}}^{{\text{opt}}}}$ \cite{El14,Bogale14,Lee12}
 \begin{equation}
\begin{gathered}
  \left( {{\mathbf{B}}_{{\text{RF}}}^{{\text{opt}}},{\mathbf{B}}_{{\text{BB}}}^{{\text{opt}}}} \right) = \mathop {\arg \;\min }\limits_{{{\mathbf{B}}_{{\text{RF}}}},\;\;{{\mathbf{B}}_{{\text{BB}}}}} \;{\left\| {{{\mathbf{B}}^{{\text{opt}}}} - {{\mathbf{B}}_{{\text{RF}}}}{{\mathbf{B}}_{{\text{BB}}}}} \right\|_F} \hfill \\
  \;\;\;\;\;\;\;\;\;s.t.\;\;\;{\left| {{{\left[ {{{\mathbf{B}}_{{\text{RF}}}}} \right]}_{i,j}}} \right|^2} = \frac{1}{{{N_{{\text{Tx}}}}}} \hfill \\
\end{gathered}.
	\label{eq17}
	\tag{17}
	\end{equation}

Considering the non-convex constraint ${\left| {{{\left[ {{{\mathbf{B}}_{{\text{RF}}}}} \right]}_{i,j}}} \right|^2} = \frac{1}{{{N_{{\text{Tx}}}}}}$ \cite{Boyd04}, it is not tractable to analytically solve the optimization problem in (\ref{eq17}). Based on the millimeter wave channel in (\ref{eq3}), the entries of BS antenna array steering matrix ${\mathbf{U}} = \left[ {{\mathbf{u}}\left( {{\psi _1},{\vartheta _1}} \right),...,{\mathbf{u}}\left( {{\psi _i},{\vartheta _i}} \right),...,{\mathbf{u}}\left( {{\psi _{{N_{{\text{ray}}}}}},{\vartheta _{{N_{{\text{ray}}}}}}} \right)} \right] \in {\mathbb{C}^{{N_{{\text{Tx}}}} \times {N_{{\text{ray}}}}}}$ are constant-amplitude which can be implemented by phase shifters in BS RF circuits. Meanwhile, as pointed out in \cite{El14}, the columns vectors of steering matrix ${\mathbf{U}}$ are independent from each other in millimeter wave channels. Moreover, ${\mathbf{U}} \in {\mathbb{C}^{{N_{{\text{Tx}}}} \times {N_{{\text{ray}}}}}}$ and ${{\mathbf{B}}_{{\text{RF}}}} \in {\mathbb{C}^{{N_{{\text{Tx}}}} \times {N_{{\text{RF}}}}}}$ have the same row numbers. To simplify the engineering application, ${N_{{\text{RF}}}}$ column vectors are selected from ${\mathbf{U}}$ to form the column vectors of ${{\mathbf{B}}_{{\text{RF}}}}$. The detailed vector selection method is described in the EEHP-B algorithm. Furthermore, the baseband precoding matrix ${{\mathbf{B}}_{{\text{BB}}}}$ is optimized by approaching ${{\mathbf{B}}_{{\text{RF}}}}{{\mathbf{B}}_{{\text{BB}}}}$ to ${{\mathbf{B}}^{{\text{opt}}}}$. As a consequence, the optimization problem in (\ref{eq17}) is transformed as follows
 \begin{equation}
 \begin{gathered}
  {\mathbf{\overset{\lower0.5em\hbox{$\smash{\scriptscriptstyle\frown}$}}{B} }}_{{\text{BB}}}^{{\text{opt}}} = \mathop {arg\;\min }\limits_{{{{\mathbf{\bar B}}}_{{\text{BB}}}}} \;{\left\| {{{\mathbf{B}}^{{\text{opt}}}} - {\mathbf{U}}{{{\mathbf{\overset{\lower0.5em\hbox{$\smash{\scriptscriptstyle\frown}$}}{B} }}}_{{\text{BB}}}}} \right\|_F} \hfill \\
  \;\;\;\;\;\;\;\;\;s.t.\;\;\;{\left\| {{\text{diag}}\left( {{{{\mathbf{\overset{\lower0.5em\hbox{$\smash{\scriptscriptstyle\frown}$}}{B} }}}_{{\text{BB}}}}{\mathbf{\overset{\lower0.5em\hbox{$\smash{\scriptscriptstyle\frown}$}}{B} }}_{{\text{BB}}}^H} \right)} \right\|_0} = {N_{{\text{RF}}}} \hfill \\
  \;\;\;\;\;\;\;\;\;\;\;\;\;\;\;{\left\| {{\mathbf{U}}{{{\mathbf{\overset{\lower0.5em\hbox{$\smash{\scriptscriptstyle\frown}$}}{B} }}}_{{\text{BB}}}}} \right\|^2} = {\left\| {{{\mathbf{B}}^{{\text{opt}}}}} \right\|^2} \hfill \\
\end{gathered} ,  	
	\label{eq18}
	\tag{18}
	\end{equation}
where ${{\mathbf{\overset{\lower0.5em\hbox{$\smash{\scriptscriptstyle\frown}$}}{B} }}_{{\text{BB}}}} \in {\mathbb{C}^{{N_{{\text{ray}}}} \times K}}$ is a digital precoding matrix constrained by ${\left\| {{\text{diag}}\left( {{{{\mathbf{\overset{\lower0.5em\hbox{$\smash{\scriptscriptstyle\frown}$}}{B} }}}_{{\text{BB}}}}{\mathbf{\overset{\lower0.5em\hbox{$\smash{\scriptscriptstyle\frown}$}}{B} }}_{{\text{BB}}}^H} \right)} \right\|_0} = {N_{{\text{RF}}}}$, and ${{\mathbf{\overset{\lower0.5em\hbox{$\smash{\scriptscriptstyle\frown}$}}{B} }}_{{\text{BB}}}}$ has ${N_{{\text{RF}}}}$ non-zero rows; ${\left\| {{\mathbf{U}}{{{\mathbf{\overset{\lower0.5em\hbox{$\smash{\scriptscriptstyle\frown}$}}{B} }}}_{{\text{BB}}}}} \right\|^2} = {\left\| {{{\mathbf{B}}^{{\text{opt}}}}} \right\|^2}$ is the transmit power constraint. As a result, the energy efficient hybrid precoding-B (EEHP-B) algorithm is developed for the optimization problem in (\ref{eq18}) as follow.

\begin{algorithm}[H] 
\caption{\textbf{Energy Efficient Hybrid Precoding-B (EEHP-B) algorithm}.} 
\label{alg2} 
\begin{algorithmic}
\STATE \textbf{Begin:} \begin{enumerate}
    \item Preset ${{\mathbf{B}}_{{\text{RF}}}}$ as an ${N_{{\text{Tx}}}} \times {N_{{\text{RF}}}}$ empty matrix, and set ${{\mathbf{B}}_{{\text{temp}}}} = {{\mathbf{B}}^{{\text{opt}}}}$;

    \item \textbf{For} $i = 1:1:{N_{{\text{RF}}}}$

    \ \ ${\mathbf{\Delta }} = {{\mathbf{U}}^H}{{\mathbf{B}}_{{\text{temp}}}}$;

    \ \ $v = \arg \mathop {max}\limits_{v = 1,...,{N_{{\text{ray}}}}} {\left[ {{\mathbf{\Delta }}{{\mathbf{\Delta }}^H}} \right]_{v,v}}$;

    \ \ ${{\mathbf{B}}_{{\text{RF}}}} = \left[ {{{\mathbf{B}}_{{\text{RF}}}}\left| {\;{{\left[ {\mathbf{U}} \right]}_{\;:\;,\;v}}} \right.} \right]$;

    \ \ ${{\mathbf{B}}_{{\text{BB,temp}}}} = {\left( {{\mathbf{B}}_{{\text{RF}}}^H{{\mathbf{B}}_{{\text{RF}}}}} \right)^{ - 1}}{\mathbf{B}}_{{\text{RF}}}^H{{\mathbf{B}}^{{\text{opt}}}}$;

    \ \ ${{\mathbf{B}}_{{\text{temp}}}} = \frac{{{{\mathbf{B}}^{{\text{opt}}}} - {{\mathbf{B}}_{{\text{RF}}}}{{\mathbf{B}}_{{\text{BB,temp}}}}}}{{{{\left\| {{{\mathbf{B}}^{{\text{opt}}}} - {{\mathbf{B}}_{{\text{RF}}}}{{\mathbf{B}}_{{\text{BB,temp}}}}} \right\|}_F}}}$;


	\textbf{End for}

    \item ${\mathbf{B}}_{{\text{RF}}}^{{\text{opt}}}$ is solved by the step 2. The optimal baseband precoding matrix is calculated by ${\mathbf{B}}_{{\text{BB}}}^{{\text{opt}}} = {\left\| {{{\mathbf{B}}^{{\text{opt}}}}} \right\|_F}\frac{{{{\mathbf{B}}_{{\text{BB,temp}}}}}}{{{{\left\| {{\mathbf{B}}_{{\text{RF}}}^{{\text{opt}}}{{\mathbf{B}}_{{\text{BB,temp}}}}} \right\|}_F}}}$.

                       \end{enumerate}
\STATE \textbf{end Begin}
\end{algorithmic}
\end{algorithm}

Since the EEHP-B algorithm has the specified cycle number, i.e., ${N_{{\text{RF}}}}$, the EEHP-B algorithm is guaranteed to converge.

When the optimal hybrid precoding matrices ${\mathbf{B}}_{{\text{RF}}}^{{\text{opt}}}$ and ${\mathbf{B}}_{{\text{BB}}}^{{\text{opt}}}$ are submitted into (\ref{eq8}), an optimal energy efficiency ${\eta ^{{\text{opt}}}}$ can be obtained. Based on the EEHP-B algorithm, the value of ${\eta ^{{\text{opt}}}}$ approaches to the value of ${\bar \eta ^{{\text{opt}}}}$, i.e., ${\eta ^{{\text{opt}}}} \leqslant {\bar \eta ^{{\text{opt}}}}$. Considering ${\bar \eta ^{{\text{opt}}}}$ is a locally optimal solution for the energy efficiency optimization in (9), ${\eta ^{{\text{opt}}}}$ is also a locally optimal solution for the energy efficiency optimization in (\ref{eq8}).

\subsection{number of RF chains Optimization}

Based on EEHP-A and EEHP-B algorithms, a local optimization solution ${\eta ^{{\text{opt}}}}$ with optimized hybrid precoding matrices ${\mathbf{B}}_{{\text{RF}}}^{{\text{opt}}}$ and ${\mathbf{B}}_{{\text{BB}}}^{{\text{opt}}}$ is available for the energy efficiency optimization in (\ref{eq8}). However, the number of RF chains is fixed for the solution ${\eta ^{{\text{opt}}}}$. To maximize the energy efficiency, the number of RF chains is further optimized based on the locally optimal solution ${\eta ^{{\text{opt}}}}$.

By analyzing (9) and (\ref{eq17}), the number of RF chains ${N_{{\text{RF}}}}$ is related not only to the size of the optimized hybrid precoding matrices ${\mathbf{B}}_{{\text{RF}}}^{{\text{opt}}}$ and ${\mathbf{B}}_{{\text{BB}}}^{{\text{opt}}}$ but also to the entries of ${\mathbf{B}}_{{\text{RF}}}^{{\text{opt}}}$ and ${\mathbf{B}}_{{\text{BB}}}^{{\text{opt}}}$. Therefore, it is difficulty to derive an analytical solution for the optimal number of RF chains. However, the number of RF chains ${N_{{\text{RF}}}}$ is an positive integer and is limited in the specific range $\left[ {K,{N_{{\text{Tx}}}}} \right]$. In this case, we can utilize the ergodic searching method to find the optimal number of RF chains maximizing the energy efficiency in (\ref{eq8}). Therefore, the EEHP algorithm is developed to achieve the global optimization solution for the energy efficiency optimization in (\ref{eq8}).

\begin{algorithm}[H] 
\caption{\textbf{Energy Efficient Hybrid Precoding (EEHP) algorithm}.} 
\label{alg3} 
\begin{algorithmic}
\STATE \textbf{Begin:}
\begin{enumerate}
    \item \textbf{For} ${N_{{\text{RF}}}} = K:1:{N_{{\text{Tx}}}}$ (search all the possible values of ${N_{{\text{RF}}}}$ from $K$ to ${N_{{\text{Tx}}}}$)

    \ \ For a certain value of ${N_{{\text{RF}}}}$, calculate ${{\mathbf{B}}^{{\text{opt}}}}\left( {{N_{{\text{RF}}}}} \right)$ according to the EEHP-A algorithm;

    \ \  Based on ${{\mathbf{B}}^{{\text{opt}}}}\left( {{N_{{\text{RF}}}}} \right)$ and ${N_{{\text{RF}}}}$, calculate ${\mathbf{B}}_{\text{RF}}^{{\text{opt}}}\left( {{N_{{\text{RF}}}}} \right)$ and ${\mathbf{B}}_{\text{BB}}^{{\text{opt}}}\left( {{N_{{\text{RF}}}}} \right)$ according to the \ \ \ \ \ \ \ \ \ \ \ \ \ \ \ \ \ \ \ \ \ \ \ \ \ \ \ \ \ \ \ \ \ \ \ \ \ \ \ \ \ \ \ \ \ \ \ \ \ \ \ \ \ \ \ \ \ \ \ \ \ \ \ \ \ \ \ \ \ \ \ \ \ \ \ \ \ \ \ \ \ \ \ \ \ \ \ \ \ \ \ \ \ \ \ \ \ \ \ \ \ \ \ \ \ \ \ \ \ \ \ \ \ \ \ \ \ \ \ \ \ \ \ \ \ \ \ \ \ \ \ \ \ \ \ \ \ \ \ \ \ \ \ \ ${\kern 1pt} {\kern 1pt} {\kern 1pt} {\kern 1pt} {\kern 1pt} {\kern 1pt} {\kern 1pt} {\kern 1pt} {\kern 1pt} {\kern 1pt} $EEHP-B algorithm;

    \ \  Calculate ${\eta ^{{\text{opt}}}}\left( {{N_{{\text{RF}}}}} \right)$ with ${\mathbf{B}}_{\text{RF}}^{{\text{opt}}}\left( {{N_{{\text{RF}}}}} \right)$ and ${\mathbf{B}}_{\text{BB}}^{{\text{opt}}}\left( {{N_{{\text{RF}}}}} \right)$;


	\textbf{End for}

    \item Find the optimal number of RF chains $N_{{\text{RF}}}^{{\text{opt}}}$ maximizing the energy efficiency;

    \item Configure the global optimal hybrid precoding matrices as ${\mathbf{B}}_{{\text{RF}}}^{{\text{opt}}}\left( {N_{{\text{RF}}}^{{\text{opt}}}} \right)$ and ${\mathbf{B}}_{{\text{BB}}}^{{\text{opt}}}\left( {N_{{\text{RF}}}^{{\text{opt}}}} \right)$.

                       \end{enumerate}
\STATE \textbf{end Begin}
\end{algorithmic}
\end{algorithm}

Based on the EEHP algorithm, the global maximum energy efficiency $\eta _{{\text{global}}}^{{\text{opt}}}$ is achieved by configuring the number of RF chains $N_{{\text{RF}}}^{{\text{opt}}}$, the RF chain precoding matrix ${\mathbf{B}}_{{\text{RF}}}^{{\text{opt}}}\left( {N_{{\text{RF}}}^{{\text{opt}}}} \right)$ and the baseband precoding matrix ${\mathbf{B}}_{{\text{BB}}}^{{\text{opt}}}\left( {N_{{\text{RF}}}^{{\text{opt}}}} \right)$. Considering $\eta _{{\text{global}}}^{{\text{opt}}} \leqslant {\bar \eta ^{{\text{opt}}}} \leqslant {\bar \eta ^{\max }}$, an suboptimal energy efficiency optimization solution is found by the EEHP algorithm in this section.

{Furthermore, according to the computational complexity of matrix calculation and iterative algorithms in \cite{Golub96} and \cite{He14}, the computational complexity of the proposed algorithm is explained as follows: The complexity of Algorithm 1 is calculated as $O\left( {{N_{{\text{Tx}}}}K} \right) + O\left( {N_{{\text{Tx}}}^3} \right)$ floating point operations (flops); the complexity of Algorithm 2 is calculated as $O\left( {N_{{\text{Tx}}}^2K + N_{{\text{RF}}}^3 + N_{{\text{RF}}}^2{N_{{\text{Tx}}}} + {N_{{\text{Tx}}}}{N_{{\text{RF}}}}K} \right)$ flops; combining Algorithm 1 and 2, the complexity of the EEHP algorithm, i.e. Algorithm 3 is calculated as }
$O\left[ {\left( {{N_{{\text{Tx}}}} - K} \right)\left( {N_{{\text{Tx}}}^2K + N_{{\text{RF}}}^3 + N_{{\text{RF}}}^2{N_{{\text{Tx}}}}} \right.} \right.$
$\left. {\left. { + {N_{{\text{Tx}}}}{N_{{\text{RF}}}}K + N_{{\text{Tx}}}^3} \right)} \right]$ {flops.}

\section{Energy Efficient Optimization with the Minimum number of RF chains}

In general, the cost of RF chain is very high in wireless communication systems. To reduce the cost of massive MIMO systems, an energy efficient solution with the minimum number of RF chains is investigated in the following.

\subsection{Energy Efficiency Hybrid Precoding with the Minimum number of RF chains}

Based on the function of RF chains and the hybrid precoding scheme, the number of RF chains is larger than or equal to the number of baseband data streams in massive MIMO communication systems \cite{Zhang05,Pei12}.  In this paper the number of baseband data streams is assumed to be equal to the number of active UEs. Without loss of generality, the minimum number of RF chains is configured as $N_{{\text{RF}}}^{\min } = K$, where $K$ is the number of active UEs in Fig.~\ref{fig1}.

When the minimum number of RF chains is configured, the RF precoding matrix ${{\mathbf{B}}_{{\text{RF}}}}$ has a size of ${N_{{\text{Tx}}}} \times K$, which is exactly the size of the conjugate transpose of the downlink channel matrix. Therefore, the entry of RF precoding matrix ${{\mathbf{B}}_{{\text{RF}}}}$ is directly configured as \cite{Liang14}
 \begin{equation}
   {\left[ {{{\mathbf{B}}_{{\text{RF}}}}} \right]_{i,j}} = \frac{1}{{\sqrt {{N_{{\text{Tx}}}}} }}{e^{j{\theta _{i,j}}}},	
	\label{eq19}
	\tag{19}
	\end{equation}
where ${\left[ {{{\mathbf{B}}_{{\text{RF}}}}} \right]_{i,j}}$ denotes the $\left( {i,j} \right){\text{th}}$ entry of the RF precoding matrix ${{\mathbf{B}}_{{\text{RF}}}}$, and ${\theta _{i,j}}$ is the phase of the $\left( {i,j} \right){\text{th}}$ entry of the conjugate transpose of the downlink channel matrix.

When the downlink channel matrix ${{\mathbf{H}}^H}$ and the RF chain precoding matrix are combined together, an equivalent downlink channel matrix for all $K$ UEs is given by
 \begin{equation}
   	\begin{gathered}
  {\mathbf{H}}_{eq}^H = {{\mathbf{H}}^H}{{\mathbf{B}}_{{\text{RF}}}} = {\left[ {{\mathbf{B}}_{{\text{RF}}}^H{{\mathbf{h}}_1},...,{\mathbf{B}}_{{\text{RF}}}^H{{\mathbf{h}}_k},...,{\mathbf{B}}_{{\text{RF}}}^H{{\mathbf{h}}_K}} \right]^H} \hfill \\
  \;\;\;\;\;\; = {\left[ {{{\mathbf{h}}_{1,eq}},...,{{\mathbf{h}}_{k,eq}},...,{{\mathbf{h}}_{K,eq}}} \right]^H} \in {\mathbb{C}^{K \times {N_{{\text{RF}}}}}} \hfill \\
\end{gathered} .
	\label{eq20}
	\tag{20}
	\end{equation}

Substitute (\ref{eq20}) into (\ref{eq8}), the energy efficiency optimization problem is transformed as

\begin{equation}
\begin{gathered}
  {\mathbf{\tilde B}}_{{\text{BB}}}^{{\text{opt}}} = \mathop {\arg \;\max }\limits_{{{\mathbf{B}}_{{\text{BB}}}}} \;\tilde \eta  = \frac{{W\sum\limits_{k = 1}^K {{{\tilde R}_{k,eq}}} }}{{\frac{1}{\alpha }\sum\limits_{k = 1}^K {{{\left\| {{{\mathbf{B}}_{{\text{RF}}}}{{\mathbf{b}}_{{\text{BB}},k}}} \right\|}^2}}  + K{P_{{\text{RF}}}} + {P_{\text{C}}}}} \hfill \\
  \;\;\;\;\;\;\;\;\;\;\;\;\;\;\;\;\;\;\;\;\;\;\;\;\;\;\;{\text{s}}{\text{.t}}{\text{.}}\;\;\;{{\tilde R}_{k,eq}} \geqslant {\Gamma _k},\;k = 1,...,K \hfill \\
  \;\;\;\;\;\;\;\;\;\;\;\;\;\;\;\;\;\;\;\;\;\;\;\;\;\;\;\;\;\;\;\;\;\sum\limits_{k = 1}^K {{{\left\| {{{\mathbf{B}}_{{\text{RF}}}}{{\mathbf{b}}_{{\text{BB}},k}}} \right\|}^2}}  = \sum\limits_{k = 1}^K {{P_k}}  \leqslant {P_{{\text{max}}}} \hfill \\
  \;\;\;\;\;\;\;\;\;\;\;\;\;\;\;\;\;\;\;\;\;\;\;\;\;\;\;\;\;\;\;\;\;\; \hfill \\
\end{gathered},  	
	\label{eq21a}
	\tag{21a}
	\end{equation}

where
\[{\tilde R_{k,eq}} = {\log _2}\left( {1 + \frac{{{\mathbf{h}}_{k,eq}^H{{\mathbf{b}}_{{\text{BB}},k}}{\mathbf{b}}_{{\text{BB}},k}^H{{\mathbf{h}}_{k,eq}}}}{{\sum\limits_{i = 1,i \ne k}^K {{\mathbf{h}}_{k,eq}^H{{\mathbf{b}}_{{\text{BB}},i}}{\mathbf{b}}_{{\text{BB}},i}^H{{\mathbf{h}}_{k,eq}}}  + \sigma _n^2}}} \right). \tag{21b}\]

Similar to the optimization problem in (\ref{eq8}), the energy efficiency $\tilde \eta $ in (\ref{eq20}) is maximized by the optimal baseband precoding matrix ${\mathbf{\tilde B}}_{{\text{BB}}}^{{\text{opt}}}$.  Considering the objective function in (21) is non-concave, it is intractable to solve for the global optimum of $\tilde \eta $. Therefore, a local optimum solution is developed for (21) as follow.

Given that $\tilde \eta $ is a function of the baseband precoding vector of the $k{\text{th}}$ UE, i.e. $\tilde \eta \left( {{{\mathbf{b}}_{{\text{BB}},k}}} \right)$, the gradient of $\tilde \eta \left( {{{\mathbf{b}}_{{\text{BB}},k}}} \right)$ is derived by
 \begin{equation}
   	\frac{{\partial \eta \left( {{{\mathbf{b}}_{{\text{BB}},k}}} \right)}}{{\partial {{\mathbf{b}}_{{\text{BB}},k}}}} = \frac{2}{{{{\tilde P}^2}}}\left[ {{{{\mathbf{\tilde \Omega }}}_k} - {{{\mathbf{\tilde \Xi }}}_k}} \right]{{\mathbf{b}}_{{\text{BB}},k}},
	\label{eq22a}
	\tag{22a}
	\end{equation}
with [22b -- 22e].
\begin{equation}
   	{{\mathbf{\tilde \Omega }}_k} = \frac{{P{{\mathbf{h}}_{k,eq}}{\mathbf{h}}_{k,eq}^H}}{{\sum\limits_{j = 1}^K {{\mathbf{h}}_{k,eq}^H{{\mathbf{b}}_{{\text{BB}},j}}{\mathbf{b}}_{{\text{BB}},j}^H{{\mathbf{h}}_{k,eq}}}  + \sigma _n^2}},
	\label{eq22b}
	\tag{22b}
	\end{equation}

\begin{equation}
\begin{gathered}
  {{{\mathbf{\tilde \Xi }}}_k} = \frac{{\sum\limits_{i = 1}^K {{{\tilde R}_{i,eq}}} }}{{\alpha \ln 2}}{\mathbf{B}}_{{\text{RF}}}^H{{\mathbf{B}}_{{\text{RF}}}} +  \hfill \\
  \;\;\;\;\;\;\;P\sum\limits_{i = 1,i \ne k}^K {\left( {\frac{{{\mathbf{h}}_{i,eq}^H{{\mathbf{b}}_{{\text{BB}},i}}{\mathbf{b}}_{{\text{BB}},i}^H{{\mathbf{h}}_{i,eq}}}}{{{{\left( {{{\tilde \delta }_i}} \right)}^2} + {{\tilde \delta }_i}{\mathbf{h}}_{i,eq}^H{{\mathbf{b}}_{{\text{BB}},i}}{\mathbf{b}}_{{\text{BB}},i}^H{{\mathbf{h}}_{i,eq}}}}{{\mathbf{h}}_{i,eq}}{\mathbf{h}}_{i,eq}^H} \right)}  \hfill \\
\end{gathered},
\label{eq22c}
\tag{22c}
\end{equation}

	 \begin{equation}
\tilde P = \frac{1}{W}\left( {\frac{1}{\alpha }\sum\limits_{i = 1}^K {{{\left\| {{{\mathbf{B}}_{{\text{RF}}}}{{\mathbf{B}}_{{\text{BB}},i}}} \right\|}^2}}  + K{P_{{\text{RF}}}} + {P_{\text{C}}}} \right),
	\label{eq22d}
	\tag{22d}
	\end{equation}

	 \begin{equation}
   	{{\tilde{\delta }}_{i}}=\sum\limits_{j=1,j\ne i}^{K}{\mathbf{h}_{i,eq}^{H}{{\mathbf{b}}_{\text{BB},j}}\mathbf{b}_{\text{BB},j}^{H}{{\mathbf{h}}_{i,eq}}}+\sigma _{n}^{2},
	\label{eq22e}
	\tag{22e}
	\end{equation}
where ${{\tilde{\delta }}_{i}}$ is the sum of the received interference and noise power for the $i\text{th}$ UE, ${{\tilde R}_{k,eq}}$ is obtained by replacing $k$ with $i$ in (21b). Replacing the results of (10) by the results of (22), the baseband precoding matrix $\mathbf{\tilde{B}}_{\text{BB}}^{\text{opt}}$ is solved by the EEHP-A algorithm. Substituting the baseband precoding matrix $\mathbf{\tilde{B}}_{\text{BB}}^{\text{opt}}$ into the EEHP algorithm, a local optimum of $\tilde{\eta }$ is solved. To differentiate this approach from the EEHP algorithm, this algorithm is denoted as the energy efficient hybrid precoding with the minimum number of RF chains (EEHP-MRFC) algorithm.

\begin{figure}
\vspace{0.1in}
\centerline{\includegraphics[width=9cm,draft=false]{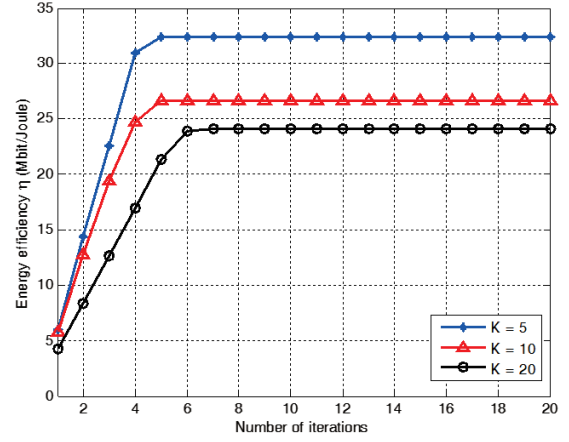}}
\caption{\small \ \ Energy efficiency of the EEHP-MRFC algorithm.}
\label{fig2}
\end{figure}

To analyze the performance of EEHP-MRFC algorithm, numerical simulation results are illustrated in Fig.~\ref{fig2}. In generally, the optimization result of EEHP-MRFC algorithm depends on the initial values of the baseband precoding matrix, i.e. $\mathbf{B}_{\text{BB}}^{\left( 0 \right)}$. Without loss of generality, $\mathbf{B}_{\text{BB}}^{\left( 0 \right)}$ can be configured to have equal entries as \cite{Jiang13}
 \begin{equation}
   	\mathbf{B}_{\text{BB}}^{\left( 0 \right)}=\sqrt{\frac{{{P}_{\max }}}{K}}{{\mathbf{1}}^{K\times K}},
	\label{eq23}
	\tag{23}
	\end{equation}
where ${{\mathbf{1}}^{K\times K}}$ denotes the $K\times K$ matrix whose entries are equal to 1, the coefficient $\sqrt{\frac{{{P}_{\max }}}{K}}$ is due to the maximum transmit power constraint in the BS. The detailed simulation parameters are list in Table I at the top of this page. Based on the results in Fig.~\ref{fig2}, the EEHP-MRFC algorithm converges after a limited iteration number. Besides, the convergence rate decreases the number of UEs increases. Moreover, the converged value of the energy efficiency also decreases with the increase of the number of UEs. This result indicates that the energy efficiency of massive MIMO system is inversely proportional to the number of active UEs.
\begin{table*}[!hbt]
\centering
\caption{\ Simulation parameters of EEHP algorithm \cite{Emil15,Sulyman14,Akdeniz14,Pei12}}
\begin{tabular}{l|l}
\hline \textbf{Parameter} & \textbf{Value} \\
\hline
Maximum transmit power ${{P}_{\max }}$ & 33 dBm \\
Minimum spectrum efficiency for each UE $\Gamma_k$ & 3 bit/s/Hz \\
Power consumed by each RF chain ${{P}_{\text{RF}}}$ & 48 mW \\
Power consumed by other parts of the BS ${{P}_{\text{C}}}$ & 20W \\
The number of BS antennas ${{N}_{\text{Tx}}}$ & 200 \\
Power amplifier efficiency $\alpha $ & 0.38\\
Noise power spectral density & -174 dBm/Hz \\
Carrier frequency & 28 GHz \\
Bandwidth & 20 MHz \\
Cell radius & 200 m \\
Minimum distance between the UE and the BS & 10 m \\
Path loss exponent $\gamma $ & 4.6 \\
The number of multipaths ${{N}_{\text{ray}}}$ & 30\\
Azimuth ${{\psi }_{i}}$ and elevation angle ${{\vartheta }_{i}}$ & Uniformly distributed within $\left[ 0,2\pi  \right]$\\

\hline
\end{tabular}
\label{tab1}

\end{table*}

\subsection{Energy Efficiency Optimization Considering the Numbers of Antennas and UEs}

In practical engineering applications, the optimal number of transmit antennas and the optimal number of UEs can be performed by the resource management and the user schedule schemes for maximizing the energy efficiency of 5G wireless communication systems. However, the EEHP algorithm can not directly derive the analytical number of transmit antennas and UEs for maximizing the energy efficiency of 5G wireless communication systems. Therefore, we try to derive the optimal number of transmit antennas and UEs for maximizing the energy efficiency of 5G massive MIMO communication systems. {To simplify the derivation, a specified scenario with rich scattering and multipaths propagation in a millimeter wave wireless channels is considered for the ergodic capacity calculation in this paper. Based on measurement results of millimeter wave channels in \cite{Raghavan11,Kalivas95 ,Yong07}, the Rayleigh fading model can be adopted to describe the considered millimeter wave wireless channels.}


Assume that the transmitter, i.e., the BS has the perfect CSI. When the minimum number of RF chains is configured and the ZF precoding is adopted in the system model \cite{Vu07}, based on (\ref{eq19}) and (\ref{eq20}), the baseband precoding matrix is given by
 \begin{equation}
  {{\mathbf{B}}_{\text{BB}}}={{\mathbf{H}}_{eq}}{{\left( \mathbf{H}_{eq}^{H}{{\mathbf{H}}_{eq}} \right)}^{-1}}\mathbf{D} ,
	\label{eq24}
	\tag{24}
	\end{equation}
where $\mathbf{D}$ is a $K\times K$ diagonal matrix, which aims to normalize ${{\mathbf{B}}_{\text{BB}}}$. Assume that ${{P}_{\text{out}}}$ is the total BS downlink transmit power consumed by $K$ active UEs. To simplify the derivation, the equal power allocation scheme is assumed to be adopted for all $K$ active UEs.
 Substituting (\ref{eq24}) into (\ref{eq5}), the normalized link capacity, i.e. the link spectral efficiency of $\text{U}{{\text{E}}_{k}}$ is expressed as \cite{Ngo13}
 \begin{equation}
 {R_{k,{\text{ZF}}}} = {\log _2}\left( {1 + \frac{{{P_{{\text{out}}}}}}{{K{{\left[ {{{\left( {{{\mathbf{H}}_{eq}}{\mathbf{H}}_{eq}^H} \right)}^{ - 1}}} \right]}_{k,k}}}}} \right).
	\label{eq25}
	\tag{25}
	\end{equation}

Based on Jensen inequality, the upper-bound of the ergodic link capacity of $\text{U}{{\text{E}}_{k}}$ is derived by
\begin{equation}
\mathbb{E}\left( {{R_{k,{\text{ZF}}}}} \right) \leqslant {\hat R_{k,{\text{ZF}}}} = {\log _2}\left[ {1 + \mathbb{E}\left( {\frac{{{P_{{\text{out}}}}}}{{K{{\left[ {{{\left( {{{\mathbf{H}}_{eq}}{\mathbf{H}}_{eq}^H} \right)}^{ - 1}}} \right]}_{k,k}}}}} \right)} \right],
	\label{eq26}
	\tag{26}
	\end{equation}
{where $\mathbb{E}\left( \centerdot  \right)$ is the expectation operation taken over the Rayleigh fading channel $\mathbf{H}$ within ${{\mathbf{H}}_{eq}} = {\mathbf{B}}_{{\text{RF}}}^H{\mathbf{H}}$.} Diagonal and off-diagonal entries of ${{\mathbf{H}}_{eq}}$ are given by
 \begin{equation}
{\left[ {{{\mathbf{H}}_{eq}}} \right]_{k,k}} = {\mathbf{h}}_k^H{{\mathbf{b}}_{{\text{RF}},k}} = \frac{1}{{\sqrt {{N_{{\text{Tx}}}}} }}\sum\limits_{i = 1}^{{N_{Tx}}} {\left| {{{\left[ {\mathbf{H}} \right]}_{i,k}}} \right|},  	
	\label{eq27a}
	\tag{27a}
	\end{equation}

 \begin{equation}
{\left[ {{{\mathbf{H}}_{eq}}} \right]_{j,k}} = {\mathbf{h}}_j^H{{\mathbf{b}}_{{\text{RF}},k}} = \frac{1}{{\sqrt {{N_{{\text{Tx}}}}} }}\sum\limits_{i = 1}^{{N_{Tx}}} {{{\left[ {\mathbf{H}} \right]}_{i,j}}{e^{j{\theta _{i,k}}}}},	
	\label{eq27b}
	\tag{27b}
	\end{equation}
where ${\mathbf{b}}_{\text{RF},k}$ is the $k$th column of ${\mathbf{B}}_{\text{RF}}$.
With Rayleigh fading channel considered and based on results in \cite{Liang14}, the diagonal entry of ${{\mathbf{H}}_{eq}}$ is governed by a normal distribution, i.e., ${{\left[ {{\mathbf{H}}_{eq}} \right]}_{k,k}}\sim \mathcal{N}\left( \frac{\pi \sqrt{{{N}_{\text{Tx}}}}}{2},1-\frac{\pi }{4} \right)$ and the off-diagonal entry of ${{\mathbf{H}}_{eq}}$ is governed by a standard normal distribution, i.e., ${{\left[ {{\mathbf{H}}_{eq}} \right]}_{j,k}}\sim \mathcal{C}\mathcal{N}\left( 0,1 \right)$. {Considering massive MIMO antennas are equipped at the BS, without loss of generality, the number of BS antennas is assumed to be larger than or equal to 100, i.e. ${{N}_{\text{Tx}}} \geqslant 100$. In this case the expected value of diagonal entries is much larger than the expected value of off-diagonal entries in ${{\mathbf{H}}_{eq}}$.} Therefore, the expected value of the off-diagonal entries can be set to zero in ${{\mathbf{H}}_{eq}}$, and ${{\mathbf{H}}_{eq}}$ is approximated as a diagonal matrix \cite{Liang14}. Based on (26), the upper-bound of the ergodic link capacity of $\text{U}{{\text{E}}_{k}}$ is approximated as
\begin{equation}
\begin{gathered}
  {{\hat R}_{k,{\text{ZF}}}} \approx {\log _2}\left[ {1 + \mathbb{E}\left( {{P_k}\left[ {{{\mathbf{H}}_{eq}}} \right]_{k,k}^2} \right)} \right] \hfill \\
  \;\;\;\;\;\;\; = {\log _2}\left[ {1 + \mathbb{E}\left( {{P_k}\left( {\frac{{{N_{{\text{Tx}}}}{\pi ^2} - \pi  + 4}}{4}} \right)} \right)} \right] \hfill \\
\end{gathered},
\label{eq28}
\tag{28}
\end{equation}

When the ergodic link capacity is replaced by the upper-bound of the ergodic link capacity, the upper-bound of the BS energy efficiency is derived by
 \begin{equation}
  {{\hat{\eta }}_{\text{ZF}}}=\frac{K{{\log }_{2}}\left( 1+\frac{{{P}_{\text{out}}}}{K}\left( \frac{{{N}_{\text{Tx}}}{{\pi }^{2}}-\pi +4}{4} \right) \right)}{\frac{1}{\alpha }{{P}_{\text{out}}}+K\left( {{P}_{\text{RF}}}+{{P}_{\text{BB}}} \right)+{{{{P}'}}_{\text{C}}}}, 	
	\label{eq29}
	\tag{29}
	\end{equation}
where ${{P}_{\text{BB}}}$ is the power consumed by the baseband processing for the baseband data stream, ${{{P}'}_{\text{C}}}$ is the fixed BS power consumption without the power consumed for the downlink transmit, the RF chains and the baseband processing.

For the upper-bound of the BS energy efficiency in (\ref{eq29}), the following proposition is given.

\textbf{Proposition: }Considering the impact of the number of UEs and transmit antennas on the BS energy efficiency, a function $\mathbb{G}\left( K,{{N}_{Tx}} \right)$ is formed as (30)
 \begin{equation}
\begin{gathered}
  \mathbb{G}\left( {K,{N_{Tx}}} \right) = \left[ {\frac{{{P_{{\text{out}}}}\left( {{P_{{\text{RF}}}} + {P_{{\text{BB}}}}} \right)\left( {{N_{{\text{Tx}}}}{\pi ^2} - \pi  + 4} \right)}}{{\left( {\frac{4}{\alpha }{P_{{\text{out}}}} + 4{{P'}_{\text{C}}}} \right)ln2}} + } \right.{\kern 1pt} {\kern 1pt}  \hfill \\
  {\kern 1pt} \left. {\frac{{{P_{{\text{out}}}}\left( {{N_{{\text{Tx}}}}{\pi ^2} - \pi  + 4} \right)}}{{4K\ln 2}}} \right]\;{\text{ - }}\left[ {1 + \frac{{{P_{{\text{out}}}}\left( {{N_{{\text{Tx}}}}{\pi ^2} - \pi  + 4} \right)\;\;}}{{4K}}} \right] \times  \hfill \\
  {\kern 1pt} {\kern 1pt} {\kern 1pt} {\log _2}\left[ {1 + \frac{{{P_{{\text{out}}}}\left( {{N_{{\text{Tx}}}}{\pi ^2} - \pi  + 4} \right)\;\;}}{{4K}}} \right] \hfill \\
\end{gathered}.
	\label{eq30}
	\tag{30}
	\end{equation}

When $\mathbb{G}\left( 1,100 \right)\ge 0$, there exists a critical number of antennas $N_{\text{Tx}}^{\text{Cri}}$. The critical number of antennas $N_{\text{Tx}}^{\text{Cri}}$ is the smallest integer which is larger than or equal to the root of the equation $\mathbb{G}\left( 1,{N_{{\text{Tx}}}} \right)=0$. When ${{N}_{\text{Tx}}}\ge N_{\text{Tx}}^{\text{Cri}}$, there exists an optimal number of UEs ${{K}^{\text{opt}}}\ge 1$ which maximizes the BS energy efficiency as $\hat{\eta }_{\text{ZF}}^{\text{Max}}$. ${{K}^{\text{opt}}}$ is solved by the integer closest to the root of the equation $\mathbb{G}\left( K,{N_{{\text{Tx}}}} \right)=0$. When ${{N}_{\text{Tx}}}<N_{\text{Tx}}^{\text{Cri}}$, the maximum BS energy efficiency $\hat{\eta }_{\text{ZF}}^{\text{Max}}$ is achieved with the number of UEs $K=1$. Moreover, the BS energy efficiency ${{\hat{\eta }}_{\text{ZF}}}$ decreases with the increase of the number of UEs $K$.

When $\mathbb{G}\left( 1,100 \right)<0$, there exists an optimal number of UEs ${{K}^{\text{opt}}}$ which maximizes the BS energy efficiency as $\hat{\eta }_{\text{ZF}}^{\text{Max}}$. When the number of antennas ${{N}_{\text{Tx}}}$ is given, the optimal number of UEs ${{K}^{\text{opt}}}$ is solved by the integer closest to the root of the equation $\mathbb{G}\left( K,{N_{{\text{Tx}}}} \right)=0$.

\textbf{Proof: }The proposition is proved in Appendix.

{Note that in the above proposition and the corresponding appendix, the minimum number of antennas is set as 100 considering the massive MIMO scenario. Based on the results in \cite{Liang14}, the approximation in (28) is guaranteed to hold when the number of antennas is larger than or equal to 100.} Based on the Appendix, $\mathbb{G}\left( 1,{{N}_{\text{Tx}}} \right)$ is a monotonous decreasing function with respect to ${{N}_{\text{Tx}}}$.
Utilizing the bisection method, the critical number of antennas searching (CNAS) algorithm is developed to solve the critical number of antennas $N_{\text{Tx}}^{\text{Cri}}$.

\begin{algorithm}[H] 
\caption{\textbf{Critical Number of Antennas Searching (CNAS) algorithm}.} 
\label{alg4} 
\begin{algorithmic}
\STATE \textbf{Begin:} \begin{enumerate}
    \item Preset the initial lower bound of searching range $N_{\text{Tx}}^{\text{Low}}=100$ and the initial upper bound of searching range $N_{\text{Tx}}^{\text{High}}=1000$;

    \item Adjust the upper bound of the range until $\mathbb{G}\left( 1,N_{\text{Tx}}^{\text{High}} \right)$ is less than or equal to 0.

	\textbf{While } $\mathbb{G}\left( 1,N_{\text{Tx}}^{\text{High}} \right)>0$

	\ \ \ \ \textbf{Do } $N_{\text{Tx}}^{\text{High}}=N_{\text{Tx}}^{\text{High}}\times 2$;

	\textbf{End}

	\item Preset the median value as $N_{\text{Tx}}^{\text{Temp}}={\left( N_{\text{Tx}}^{\text{Low}}+N_{\text{Tx}}^{\text{High}} \right)}/{2}\;$.

    \item Keep searching until the median value satisfies $\mathbb{G}\left( 1,N_{\text{Tx}}^{\text{Temp}} \right)\ =0$.

    \textbf{While } $\mathbb{G}\left( 1,N_{\text{Tx}}^{\text{Temp}} \right)\ !=0$

	\ \ \ \ \textbf{If } \ $\mathbb{G}\left( 1,N_{\text{Tx}}^{\text{Temp}} \right)>0$

	\ \ \ \ \ \ \ \ \ $N_{\text{Tx}}^{\text{Low}}=N_{\text{Tx}}^{\text{Temp}}$;

	\ \ \ \ \textbf{Else }

	\ \ \ \ \ \ \ \ \ $N_{\text{Tx}}^{\text{High}}=N_{\text{Tx}}^{\text{Temp}}$;

	\ \ \ \ \textbf{End}

	\ \ \ \ $N_{\text{Tx}}^{\text{Temp}}={\left( N_{\text{Tx}}^{\text{Low}}+N_{\text{Tx}}^{\text{High}} \right)}/{2}\;$;

    \textbf{End}

    \item Configure the critical number of antennas as the smallest integer not less than the final median value, i.e.,
$N_{{\text{Tx}}}^{{\text{Cri}}} = \left\lceil {N_{{\text{Tx}}}^{{\text{Temp}}}} \right\rceil $.
                       \end{enumerate}
\STATE \textbf{end Begin}
\end{algorithmic}
\end{algorithm}

When the critical number of antennas $N_{{\text{Tx}}}^{{\text{Cri}}}$ is obtained by the CNAS algorithm, the optimal number of UEs ${K^{{\text{opt}}}}$ is solved by $\mathbb{G}\left( {K,{N_{{\text{Tx}}}}} \right) = 0$ which maximizes the BS energy efficiency $\hat \eta _{{\text{ZF}}}^{{\text{Max}}}$.
 Considering $\mathbb{G}\left( {K,{N_{{\text{Tx}}}}} \right)$ monotonously decreases with the increase of the number of UEs $K$, the UE number optimization (UENO) algorithm is developed to solve the optimal number of UEs ${K^{{\text{opt}}}}$ by the bisection method.

 {Since the above CNAS and UENO algorithm are based on the bisection method, the computational complexity of the CNAS and UENO algorithms are calculated as $O\left[ {{{\log }_2}} \right.\left( {N_{{\text{Tx}}}^{{\text{High}}}} \right.$ $\left. {\left. { - N_{{\text{Tx}}}^{{\text{Low}}}} \right)} \right]$ and $O\left[ {{{\log }_2}\left( {{K_{{\text{High}}}} - {K_{{\text{Low}}}}} \right)} \right]$, respectively \cite{Byers88}.}

\begin{algorithm}[H] 
\caption{\textbf{The UE Number Optimization (UENO) algorithm}.} 
\label{alg5} 
\begin{algorithmic}
\STATE \textbf{Begin:} \begin{enumerate}
    \item Preset the initial lower bound of the searching range ${K_{{\text{Low}}}} = 1$ and the initial upper bound of the searching range ${K_{{\text{High}}}} = 40$;

    \item Adjust the upper bound of the range until $\mathbb{G}\left( {{K_{{\text{High}}}},{N_{{\text{Tx}}}}} \right)$ is less than or equal to 0;

	\textbf{While } $\mathbb{G}\left( {{K_{{\text{High}}}},{N_{{\text{Tx}}}}} \right) > 0$

	\ \ \ \ \textbf{Do } ${K_{{\text{High}}}} = {K_{{\text{High}}}} \times 2$;

	\textbf{End}

	\item Preset the median value as ${K_{{\text{Temp}}}} = {{\left( {{K_{{\text{Low}}}} + {K_{{\text{High}}}}} \right)} \mathord{\left/
 {\vphantom {{\left( {{K_{{\text{Low}}}} + {K_{{\text{High}}}}} \right)} 2}} \right.
 \kern-\nulldelimiterspace} 2}$;

    \item Keep searching until the median value satisfies $\mathbb{G}\left( {{K_{{\text{Temp}}}},{N_{{\text{Tx}}}}} \right) = 0$;

    \textbf{While } $\mathbb{G}\left( {{K_{{\text{Temp}}}},{N_{{\text{Tx}}}}} \right)\;! = 0$

	\ \ \ \ \textbf{If } \ $\mathbb{G}\left( {{K_{{\text{Temp}}}},{N_{{\text{Tx}}}}} \right) > 0$

	\ \ \ \ \ \ \ \ \ ${K_{{\text{Low}}}} = {K_{{\text{Temp}}}}$;

	\ \ \ \ \textbf{Else }

	\ \ \ \ \ \ \ \ \ ${K_{{\text{High}}}} = {K_{{\text{Temp}}}}$;

	\ \ \ \ \textbf{End}

	\ \ \ \ ${K_{{\text{Temp}}}} = {{\left( {{K_{{\text{Low}}}} + {K_{{\text{High}}}}} \right)} \mathord{\left/
 {\vphantom {{\left( {{K_{{\text{Low}}}} + {K_{{\text{High}}}}} \right)} 2}} \right.
 \kern-\nulldelimiterspace} 2}$;

    \textbf{End}

    \item Configure the optimal number of UEs as the integer closest to the final median value, i.e.,
${K^{{\text{opt}}}} = \left\lceil {{K_{{\text{Temp}}}} - {1 \mathord{\left/
 {\vphantom {1 2}} \right.
 \kern-\nulldelimiterspace} 2}} \right\rceil $.
                       \end{enumerate}
\STATE \textbf{end Begin}
\end{algorithmic}
\end{algorithm}

\section{Simulation Results}

Energy efficiency optimization solutions with respect to the number of RF chains, transmit antennas and active UEs are simulated in the following. Without loss of generality, the number of active UEs is configured as 10. Other default parameters are listed in Table I.

{The EEHP algorithm is proposed in Section III to maximize the BS energy efficiency. To tradeoff the energy and cost efficiency of the BS RF circuits, the EEHP-MRFC algorithm is developed in Section IV. To analyze the proposed EEHP and EEHP-MRFC algorithms, the energy efficient digital precoding (EEDP) algorithm and sparse precoding algorithm are simulated for performance comparisons. In the EEDP algorithm, the energy efficiency of 5G wireless communication systems is obtained by substituting the optimal digital precoding vectors ${\mathbf{b}}_k^{opt}$ derived from the EEHP-A algorithm into (9). The sparse precoding algorithm in \cite{El14} is also simulated, which implements a spectrum efficiency maximization hybrid precoding scheme.}

\begin{figure}
\vspace{0.1in}
\centerline{\includegraphics[width=9cm,draft=false]{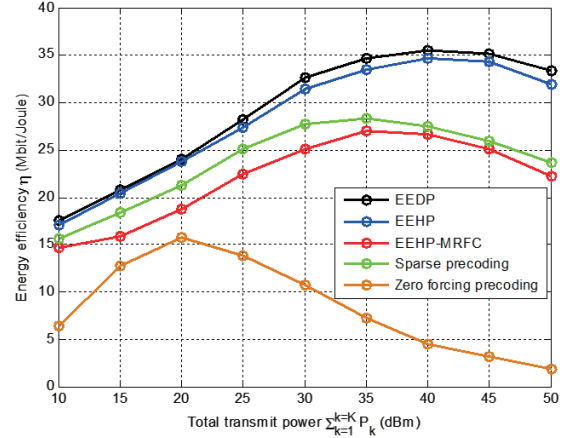}}
\caption{\small \ \ Energy efficiency with respect to the total transmit power.}
\label{fig3}
\end{figure}

\begin{figure}
\vspace{0.1in}
\centerline{\includegraphics[width=9cm,draft=false]{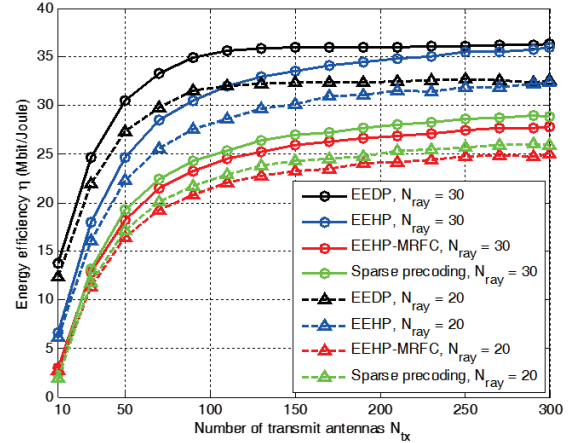}}
\caption{\small \ \ Energy efficiency with respect to the number of transmit antennas considering different numbers of multipaths.}
\label{fig4}
\end{figure}

{The two proposed algorithms along with the EEDP algorithm, the sparse precoding algorithm and the traditional ZF precoding algorithm are compared in Fig. 3. In order to better compare the energy efficiency performance of the algorithms, the total transmit power for all the UEs is used as the comparison standard \cite{Belmega11,Varma13}. For each algorithm, the energy efficiency first increases with the increasing total transmit power. When the total transmit power exceeds a given threshold, the energy efficiency starts to decrease. The variation tendencies of the curves coincide with the previously published results \cite{Belmega11,Varma13}. When the total transmit power is fixed, the EEDP algorithm has the highest energy efficiency which is consistent with the analysis result in Section III. The ZF precoding algorithm has the lowest energy efficiency because it uses the same number of RF chains as antennas, which leads to the highest power consumption for the RF chains. Besides, the sparse precoding algorithm performs more poorly than the EEHP algorithm but outperforms the EEHP-MRFC algorithm in terms of energy efficiency for 5G wireless communication systems. Although the energy efficiency of the EEHP-MRFC algorithm is not the best result, this algorithm is valuable for reducing the transmitter cost and the design complexity by minimizing the number of RF chains. In practical applications, the selection of the EEHP algorithm or the EEHP-MRFC algorithm depends on the desired tradeoff between the energy and cost efficiency for 5G wireless communication systems.}

{Fig. 4 illustrates the BS energy efficiency with respect to the number of transmit antennas considering different numbers of multipath components. The sparse precoding and EEDP algorithms are also simulated to compare with the two proposed EEHP and EEHP-MRFC algorithms. It can be seen that the BS energy efficiency increases with increasing numbers of the transmit antennas. Meanwhile, increasing the number of multipath components yields the highest BS energy efficiency when the number of transmit antennas is fixed. Decreasing energy efficiency results are yielded by the algorithms in the order EEDP, EEHP, sparse precoding and finally the EEHP-MRFC algorithm, respectively.}
\begin{figure}
\vspace{0.1in}
\centerline{\includegraphics[width=9cm,draft=false]{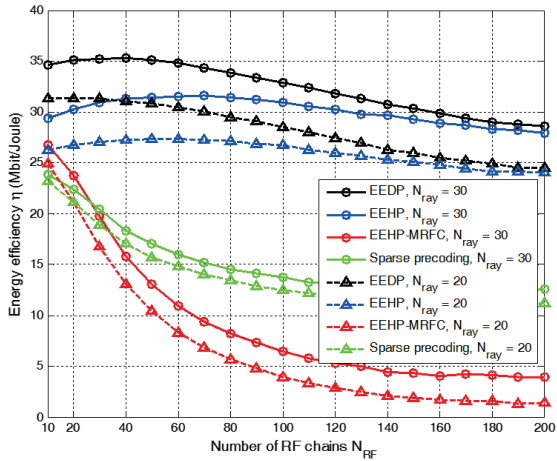}}
\caption{\small \ \ Energy efficiency with respect to the number of RF chains considering different numbers of multipaths.}
\label{fig5}
\end{figure}

{The BS energy efficiency with respect to the number of RF chains is shown in Fig. 5. As for the EEDP and EEHP algorithms, the BS energy efficiency first increases then decreases with increasing numbers of the RF chains. For the sparse precoding and EEHP-MRFC algorithms, the BS energy efficiency always decreases when increasing the number of RF chains. When the number of multipaths is fixed as 30 and the number of RF chains is less than or equal to 26, the energy efficiency of the EEHP-MRFC algorithm is larger than that of the sparse precoding algorithm. Further, when the number of RF chains is larger than 26, the energy efficiency of the EEHP-MRFC algorithm is always less than that of the sparse precoding algorithm.}

\begin{figure}
\vspace{0.1in}
\centerline{\includegraphics[width=9cm,draft=false]{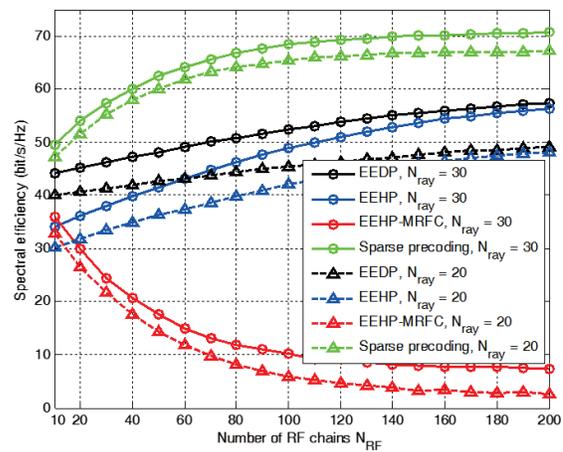}}
\caption{\small \ \ Spectral efficiency with respect to the number of RF chains considering different numbers of multipaths.}
\label{fig6}
\end{figure}

{In Fig. 6, the spectrum efficiency as a function of the number of RF chains and the number of multipaths is illustrated. It can be seen that sparse precoding algorithm has the highest spectrum efficiency. As for EEDP and EEHP algorithms, the spectrum efficiency increases with increasing the numbers of the RF chains. Considering that the number of RF chains always equals the number of UEs in the EEHP-MRFC algorithm, the spectrum efficiency of the EEHP-MRFC algorithm decreases with the increase of the number of RF chains. When the number of the RF chains is fixed, decreasing spectrum efficiency results are yielded by the algorithms in the order sparse precoding, EEDP, EEHP, and finally the EEHP-MRFC algorithm, respectively.}

\begin{figure}
\vspace{0.1in}
\centerline{\includegraphics[width=9cm,draft=false]{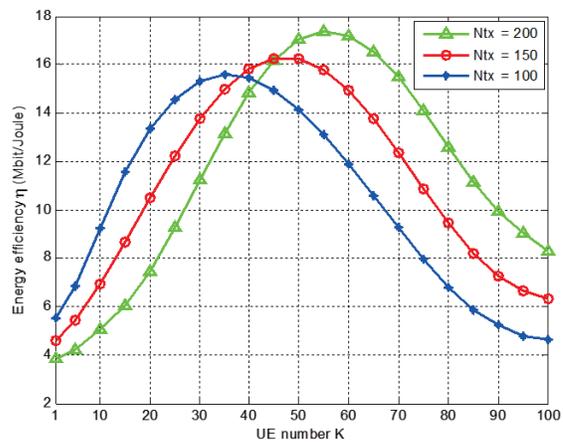}}
\caption{\small \ \ Energy efficiency with respect to the number of UEs considering ZF baseband precoding and the minimum number of RF chains.}
\label{fig7}
\end{figure}

The impact of the number of transmit antennas and the number of UEs on the BS energy efficiency is investigated in Section IV.B. Based on the CNAS and UENO algorithms, numerical simulations are shown in Fig.~\ref{fig7}. Without loss of generality, the number of transmit antennas are configured as 100, 150 and 200, respectively. The power consumed by other parts of the BS is configured as  ${P_{\text{C}}} = 20{\text{W}}$ in Fig.~\ref{fig7}. The BS energy efficiency first increases then decreases with increasing of the number of UEs. The maximums of the BS energy efficiency correspond to the optimal numbers of UEs 35, 50 and 55, respectively, which is consistent with the optimal number of UEs obtained from the proposition.

\section{Conclusion}

In this paper, the BS energy efficiency considering the energy consumption of RF chains and baseband processing is formulated as an optimization problem for 5G wireless communication systems. Considering the non-concave feature of the objective function, an available suboptimal solution is proposed by the EEHP algorithm. To tradeoff the energy and cost efficiency in RF chain circuits, the EEHP-MRFC algorithm is developed for 5G wireless communication systems. Based on the CNAS and UENO algorithms, the energy efficiency of 5G wireless communication systems can be maximized by optimizing the number of UEs and BS antennas, which is easily employed in the user scheduling and resource management schemes. Compared with the maximum energy efficiency of conventional ZF precoding algorithm, numerical results indicate that the maximum energy efficiency of the proposed EEHP and EEHP-MRFC algorithms are improved by 220\% and 171\%, respectively. Moreover, the difference between the EEHP algorithm and the EEHP-MRFC algorithm is illustrated by numerical simulation results. Furthermore, our results provide some available suboptimal energy efficiency solutions and insights into the energy and cost efficiency of RF chain circuits for 5G wireless communication systems. For the future study, we will try to investigate the energy and spectral efficiency optimization of 5G radio frequency chain systems in the multi-cell scenario.

\section*{Appendix}

\textbf{Proof to Proposition:}

To simplify the derivation, some variables are defined as follows: $z = \frac{1}{K}$, $z \in \left( {\left. {0,1} \right]} \right.$, $a = {P_{{\text{out}}}}\left( {\frac{{{N_{{\text{Tx}}}}{\pi ^2} - \pi  + 4}}{4}} \right)$, $b = \frac{1}{\alpha }{P_{{\text{out}}}} + {P'_{\text{C}}}$ and ${\text{c}} = \left( {{P_{{\text{RF}}}} + {P_{{\text{BB}}}}} \right)$. The derivative of ${\hat \eta _{{\text{ZF}}}}$ with respect to z is given by

 \begin{equation}
   	\begin{gathered}
  \frac{{{\text{d}}{{\hat \eta }_{{\text{ZF}}}}}}{{{\text{d}}z}} = \frac{{{\text{d}}\left( {\frac{{{{\log }_2}\left( {1 + az} \right)}}{{c + zb}}} \right)}}{{{\text{d}}z}} \hfill \\
  \;\;\;\;\;\;\;\;\; = \frac{{\frac{{a\left( {c + zb} \right)}}{{\left( {1 + az} \right)\ln 2}} - b{{\log }_2}\left( {1 + az} \right)}}{{{{\left( {c + zb} \right)}^2}}} \hfill \\
  \;\;\;\;\;\;\;\;\; = \frac{{\left( {\frac{{ac}}{{b\ln 2}} + \frac{a}{{\ln 2}}z} \right) - \left( {1 + az} \right){{\log }_2}\left( {1 + az} \right)}}{{\frac{1}{b}{{\left( {c + zb} \right)}^2}\left( {1 + az} \right)}} \hfill \\
\end{gathered}.
	\label{eq31}
	\tag{31}
	\end{equation}

When the numerator of (\ref{eq31}) is defined by
 \begin{equation}
  f\left( {z,a} \right) = \left( {\frac{{ac}}{{b\ln 2}} + \frac{a}{{\ln 2}}z} \right) - \left( {1 + az} \right){\log _2}\left( {1 + az} \right), 	
	\label{eq32}
	\tag{32}
	\end{equation}
(\ref{eq31}) is simply rewritten as $\frac{{{\text{d}}{{\hat \eta }_{{\text{ZF}}}}}}{{{\text{d}}z}} = \frac{{f\left( {z,a} \right)}}{{\frac{1}{b}{{\left( {c + zb} \right)}^2}\left( {1 + az} \right)}}$.

Since $\frac{{\partial f\left( {z,a} \right)}}{{\partial z}} = \frac{a}{{\ln 2}} - \left( {a\log \left( {1 + az} \right) + \frac{a}{{\ln 2}}} \right) < 0$, $f\left( {z,a} \right)$ monotonously decreases with increase of $z \in \left( {\left. {0,1} \right]} \right.$. Moreover, the limitation of $f\left( {z,a} \right)$ is derived by
 \begin{equation}
 \mathop {\lim }\limits_{z \to {0^ + }} f\left( {z,a} \right) = \frac{{ac}}{{b\ln 2}} > 0.
	\label{eq33}
	\tag{33}
	\end{equation}

If $f\left( {1,a} \right) \geqslant 0$, then $f\left( {z,a} \right) \geqslant 0$, $z \in \left( {\left. {0,1} \right]} \right.$. Furthermore, we have the result $\frac{{{\text{d}}{{\hat \eta }_{{\text{ZF}}}}}}{{{\text{d}}z}} \geqslant 0$. Therefore, ${\hat \eta _{{\text{ZF}}}}$ monotonously increases with increase of $z \in \left( {\left. {0,1} \right]} \right.$. In other words, ${\hat \eta _{{\text{ZF}}}}$ monotonously decreases with the increase of $K$. In this case, the BS energy efficiency ${\hat \eta _{{\text{ZF}}}}$ is maximized by $K = 1$. If $f\left( {1,a} \right) < 0$, there must exist an optimal value ${z^{{\text{opt}}}}$, ${z^{{\text{opt}}}} \in \left( {\left. {0,1} \right]} \right.$, which ensures $f\left( {{z^{{\text{opt}}}},a} \right) = 0$. The integer closest to ${1 \mathord{\left/
 {\vphantom {1 {{z^{{\text{opt}}}}}}} \right.
 \kern-\nulldelimiterspace} {{z^{{\text{opt}}}}}}$ is the optimal number of UEs ${K^{{\text{opt}}}}$ which maximizes the BS energy efficiency ${\hat \eta _{{\text{ZF}}}}$.

The differential of $f\left( {1,a} \right)$ with respect to $a$ is derived by

 \begin{equation}
 \frac{{\partial f\left( {1,a} \right)}}{{\partial a}} = \frac{c}{{b\ln 2}} - {\log _2}\left( {1 + a} \right).
	\label{eq34}
	\tag{34}
	\end{equation}
(\ref{eq34}) implies that $\frac{{\partial f\left( {1,a} \right)}}{{\partial a}}$ decreases with increase of $a$. When the number of transmit antenna is configured as ${N_{{\text{Tx}}}} = 1$,  i.e. $a$ is minimized as ${a_{\min }}$, the corresponding differential result is ${\left. {\frac{{\partial f\left( {1,a} \right)}}{{\partial a}}} \right|_{a = {a_{\min }}}} < 0$ considering the practical value range of ${P_{{\text{out}}}}$, ${P'_{\text{C}}}$, ${P_{{\text{RF}}}}$ and ${P_{{\text{BB}}}}$ \cite{Emil15,Tombaz11,Cui04,Yang13}. Therefore, we have the result $\frac{{\partial f\left( {1,a} \right)}}{{\partial a}} < 0$ for all available values of $a$. As a consequence, $f\left( {1,a} \right)$ monotonously decreases with increase of $a \in [{a_{\min }},\infty )$.

Substitute $z = \frac{1}{K}$, $z \in \left( {\left. {0,1} \right]} \right.$, $a = {P_{{\text{out}}}}\left( {\frac{{{N_{Tx}}{\pi ^2} - \pi  + 4}}{4}} \right)$, $b = \frac{1}{\alpha }{P_{{\text{out}}}} + {P'_{\text{C}}}$ and ${\text{c}} = \left( {{P_{{\text{RF}}}} + {P_{{\text{BB}}}}} \right)$ into (\ref{eq32}), the function $\mathbb{G}\left( {K,{N_{Tx}}} \right)$ is transformed by $f\left( {z,a} \right)$.
When $f\left( {1,{a_{\min }}} \right) < 0$, i.e., $\mathbb{G}\left( {1,100} \right) < 0$,
there exists an optimal value ${z^{{\text{opt}}}}$ which ensures $f\left( {{z^{{\text{opt}}}},a} \right) = 0$ and maximizes the BS energy efficiency ${\hat \eta _{{\text{ZF}}}}$. The integer closest to ${1 \mathord{\left/
 {\vphantom {1 {{z^{{\text{opt}}}}}}} \right.
 \kern-\nulldelimiterspace} {{z^{{\text{opt}}}}}}$ is the corresponding optimal number of UEs ${K^{{\text{opt}}}}$.

When $f\left( {1,{a_{\min }}} \right) \geqslant 0$, i.e. $\mathbb{G}\left( {1,100} \right) \geqslant 0$, there exist a critical number of antennas $N_{{\text{Tx}}}^{{\text{Cri}}}$ which ensures $f\left( {1,a} \right)\left| {_{{N_{{\text{Tx}}}} = N_{{\text{Tx}}}^{{\text{Cri}}}}} \right. = 0$. When ${N_{{\text{Tx}}}} \geqslant N_{{\text{Tx}}}^{{\text{Cri}}}$, there exists an optimal value ${z^{{\text{opt}}}}$ which ensures $f\left( {{z^{{\text{opt}}}},a} \right) = 0$ and maximizes the BS energy efficiency ${\hat \eta _{{\text{ZF}}}}$. As a consequence, the integer closest to ${1 \mathord{\left/
 {\vphantom {1 {{z^{{\text{opt}}}}}}} \right.
 \kern-\nulldelimiterspace} {{z^{{\text{opt}}}}}}$ is the corresponding optimal number of UEs ${K^{{\text{opt}}}}$. When ${N_{{\text{Tx}}}} < N_{{\text{Tx}}}^{{\text{Cri}}}$, ${\hat \eta _{{\text{ZF}}}}$ monotonously decreases with the increase of $K$. In this case, the BS energy efficiency ${\hat \eta _{{\text{ZF}}}}$ is maximized by $K = 1$.


\begin{IEEEbiography}[{\includegraphics[width=1in,height=1.25in,clip,keepaspectratio]{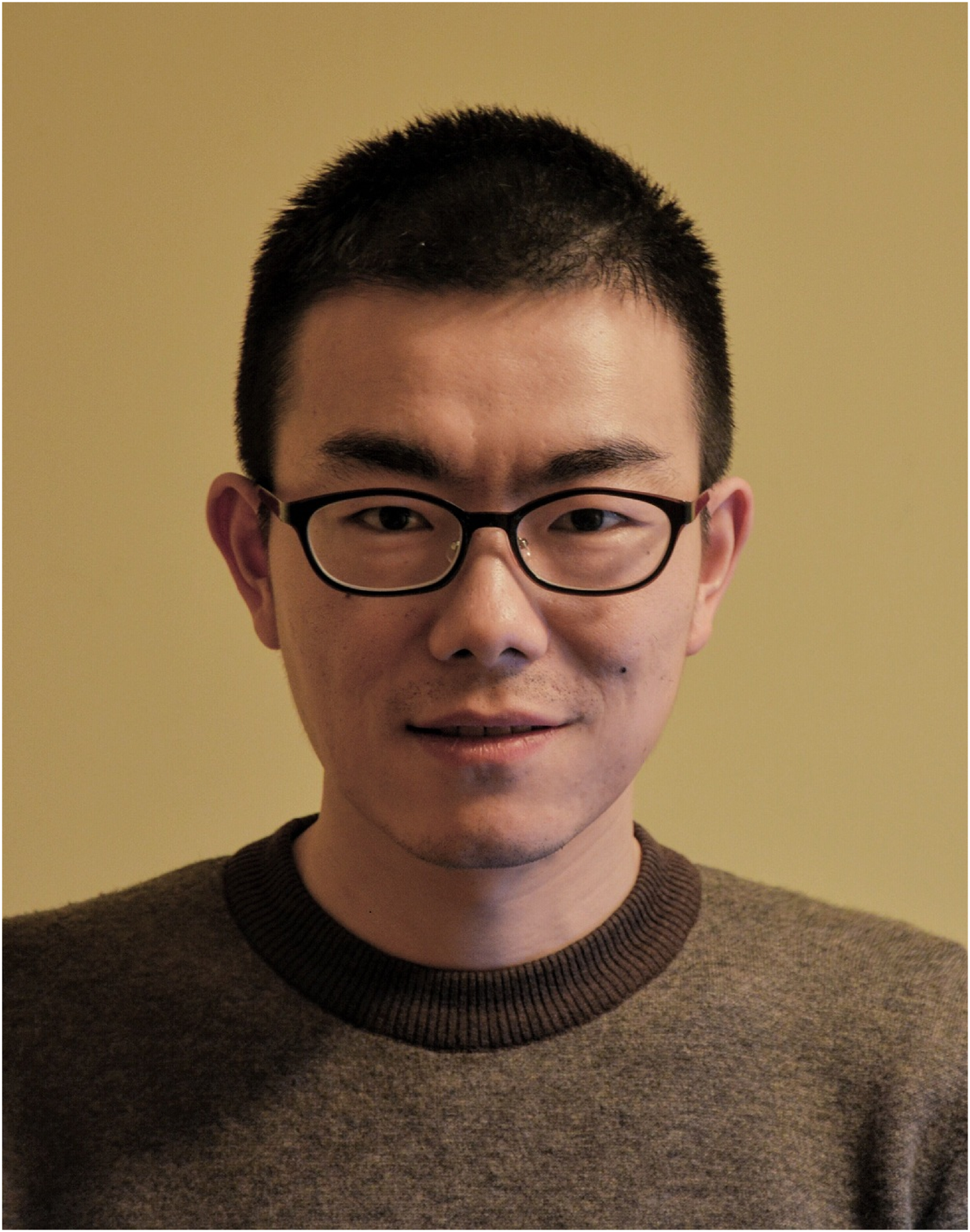}}]{Ran~Zi}
(S'14) received the B.E. degree in Communication Engineering and M.S. degree in Electronics and Communication Engineering from Huazhong University of Science and Technology (HUST), Wuhan, China in 2011 and 2013, respectively. He is currently working toward the Ph.D. degree in HUST. His research interests include MIMO systems, millimeter wave communications and multiple access technologies.
\end{IEEEbiography}

\begin{IEEEbiography}[{\includegraphics[width=1in,height=1.25in,clip,keepaspectratio]{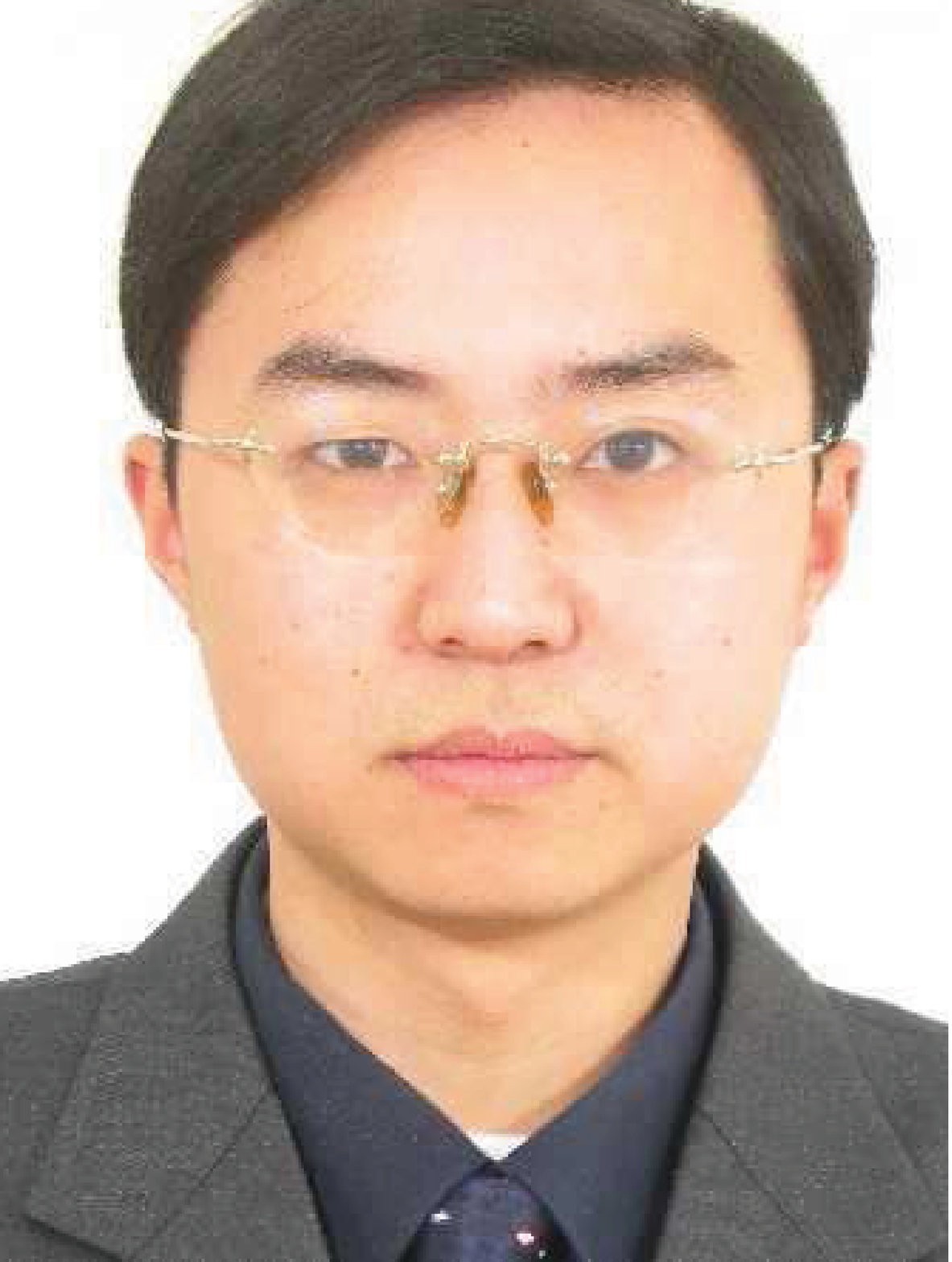}}]{Xiaohu~Ge}
(M'09-SM'11) is currently a full Professor with the School of Electronic Information and Communications at Huazhong University of Science and Technology (HUST), China. He is an adjunct professor with with the Faculty of Engineering and Information
Technology at University of Technology Sydney (UTS), Australia. He received his PhD degree in Communication and Information Engineering from HUST in 2003. He has worked at HUST since Nov. 2005. Prior to that, he worked as a researcher at Ajou University (Korea) and Politecnico Di Torino (Italy) from Jan. 2004 to Oct. 2005. He was a visiting researcher at Heriot-Watt University, Edinburgh, UK from June to August 2010. His research interests are in the area of mobile communications, traffic modeling in wireless networks, green communications, and interference modeling in wireless communications. He has published about 100 papers in refereed journals and conference proceedings and has been granted about 15 patents in China. He received the Best Paper Awards from IEEE Globecom 2010. He is leading several projects funded by NSFC, China MOST, and industries. He is taking part in several international joint projects, such as the EU FP7-PEOPLE-IRSES: project acronym WiNDOW (grant no. 318992) and project acronym CROWN (grant no. 610524).

Dr. Ge is a Senior Member of the China Institute of Communications and a member of the National Natural Science Foundation of China and the Chinese Ministry of Science and Technology Peer Review College. He has been actively involved in organizing more the ten international conferences since 2005. He served as the general Chair for the 2015 IEEE International Conference on Green Computing and Communications (IEEE GreenCom). He serves as an Associate Editor for the \textit{IEEE ACCESS}, \textit{Wireless Communications and Mobile Computing Journal (Wiley)} and \textit{the International Journal of Communication Systems (Wiley)}, etc. Moreover, he served as the guest editor for \textit{IEEE Communications Magazine} Special Issue on 5G Wireless Communication Systems.
\end{IEEEbiography}

\begin{IEEEbiography}[{\includegraphics[width=1in,height=1.25in,clip,keepaspectratio]{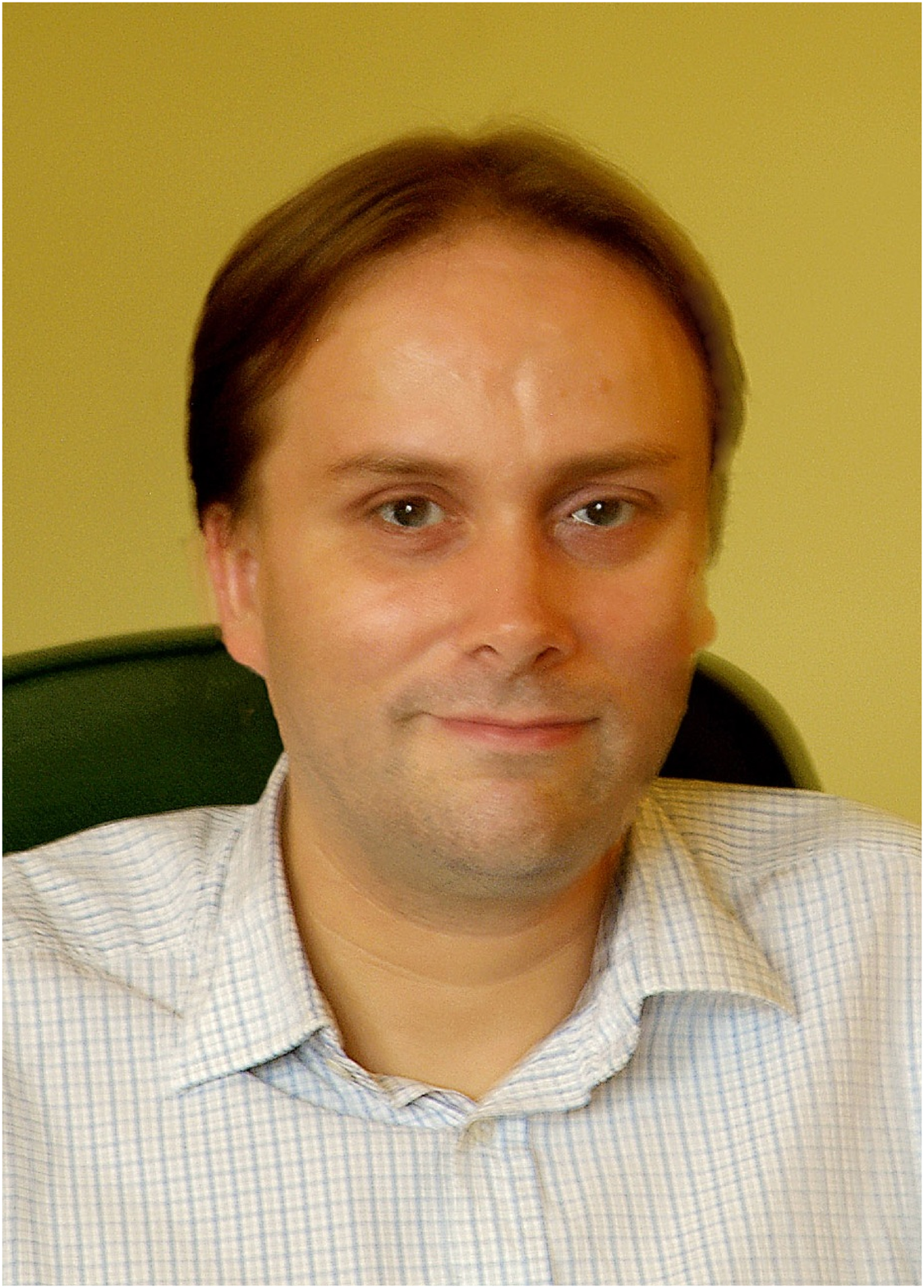}}]{John~Thompson}
(M'03) Prof. John S. Thompson is currently a Professor in Signal Processing and
Communications at the School of Engineering in the University of
Edinburgh. He specializes in antenna array processing, cooperative
communications systems and energy efficient wireless communications. He
has published in excess of three hundred papers on these topics,
including one hundred journal paper publications. He is currently the
project coordinator for the EU Marie Curie International Training Network
project ADVANTAGE, which studies how communications and power
engineering can provide  future ``smart grid'' systems). He was an elected
Member-at-Large for the Board of Governors of the IEEE Communications
Society from 2012-2014, the second largest IEEE Society. He is also a
distinguished lecturer on the topic of energy efficient communications
and smart grid for the IEEE Communications Society during 2014-2015. He
is an editor for the Green Communications and Computing Series that
appears regularly in IEEE Communications Magazine.
\end{IEEEbiography}

\begin{IEEEbiography}[{\includegraphics[width=1.4in,height=1.3in,clip,keepaspectratio]{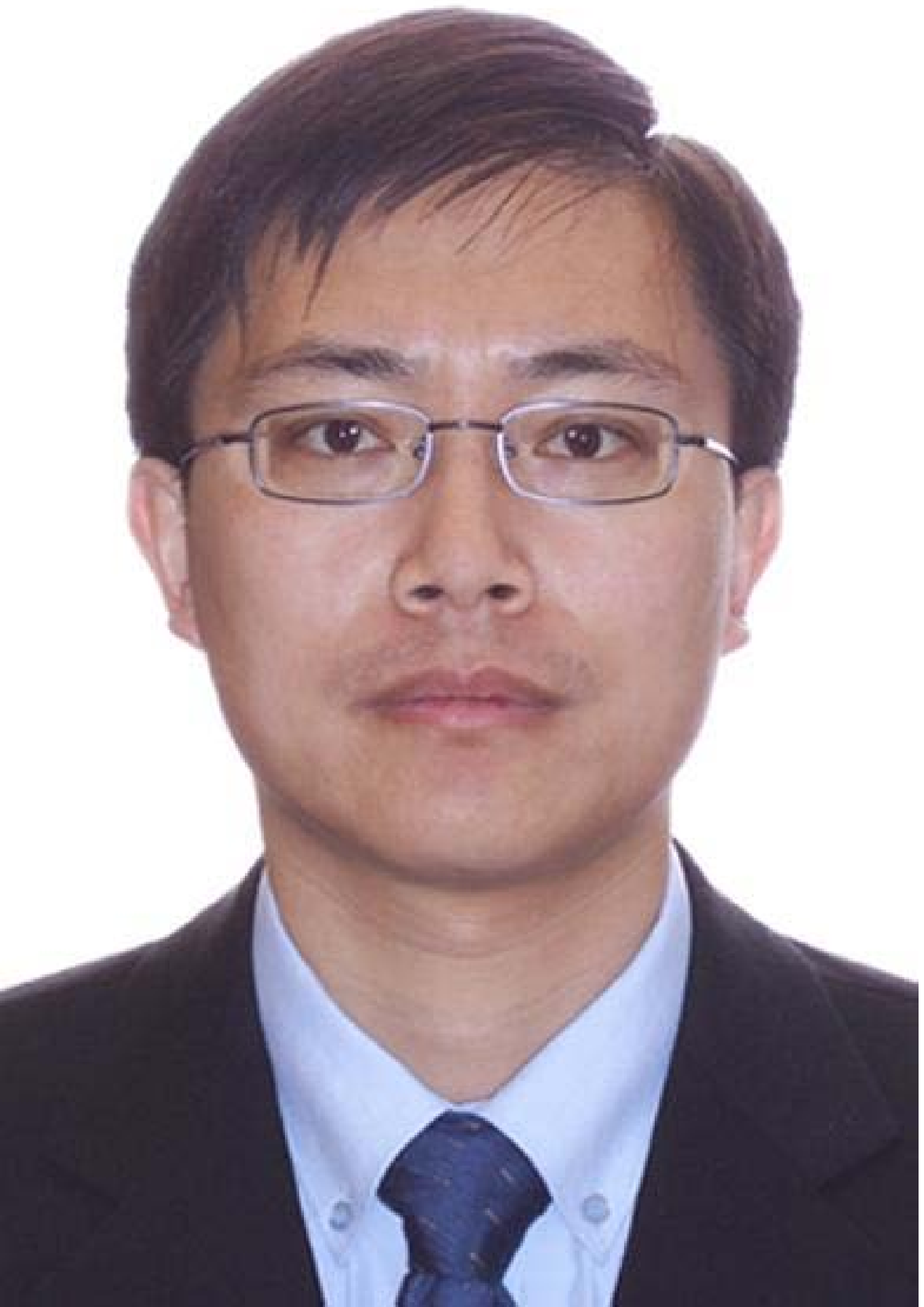}}]{Cheng-Xiang Wang}
(S'01-M'05-SM'08) received the BSc and MEng degrees in Communication and Information Systems from Shandong University, China, in 1997 and 2000, respectively, and the PhD degree in Wireless Communications from Aalborg University, Denmark, in 2004.

He has been with Heriot-Watt University, Edinburgh, U.K., since 2005, and was promoted to a Professor in wireless communications in 2011. He was a Research Fellow at the University of Agder, Grimstad, Norway, from 2001-2005, a Visiting Researcher at Siemens AG-Mobile Phones, Munich, Germany, in 2004, and a Research Assistant at Technical University of Hamburg-Harburg, Hamburg, Germany, from 2000-2001. His current research interests focus on wireless channel modelling and 5G wireless communication networks. He has edited 1 book and published over 230 papers in refereed journals and conference proceedings.

Prof. Wang served or is currently serving as an editor for 9 international journals, including IEEE Transactions on Vehicular Technology (since 2011), IEEE Transactions on Communications (since 2015), and IEEE Transactions on Wireless Communications (2007-2009). He was the leading Guest Editor for IEEE Journal on Selected Areas in Communications, Special Issue on Vehicular Communications and Networks. He served or is serving as a TPC member, TPC Chair, and General Chair for over 80 international conferences. He received the Best Paper Awards from IEEE Globecom 2010, IEEE ICCT 2011, ITST 2012, IEEE VTC 2013-Spring, and IWCMC 2015. He is a Fellow of the IET, a Fellow of the HEA, and a member of EPSRC Peer Review College.
\end{IEEEbiography}

\begin{IEEEbiography}[{\includegraphics[width=1.4in,height=1.3in,clip,keepaspectratio]{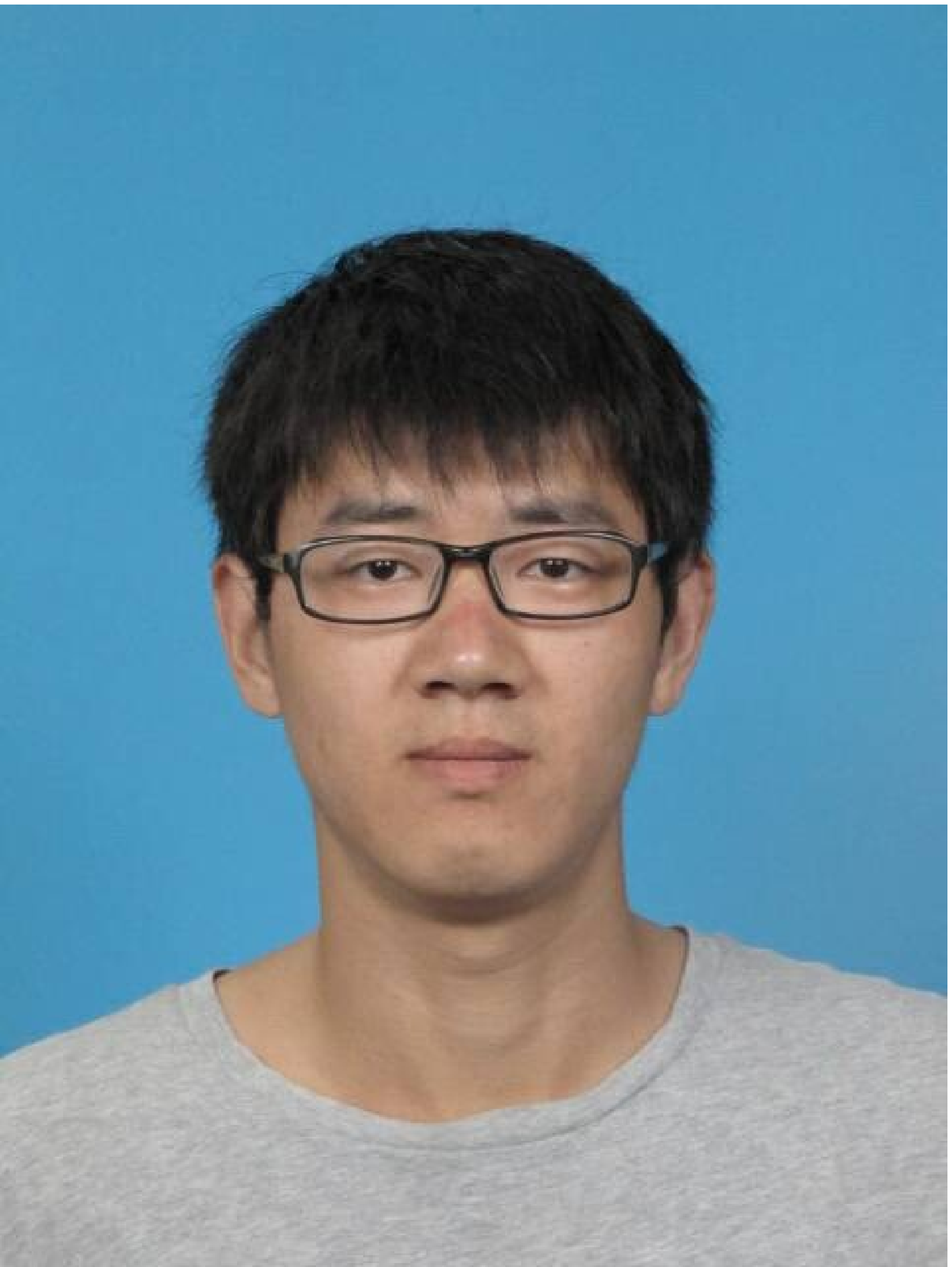}}]{Haichao Wang}
received the bachelor degree in electronic science and technology from Wuhan University of Technology, Wuhan, China, in 2013, Now he is working toward the master degree in Huazhong University of Science and Technology, Wuhan, China. His research interests include the mutual coupling effect in antenna arrays and optimization of the number of RF chains in antenna arrays.

\end{IEEEbiography}

\begin{IEEEbiography}[{\includegraphics[width=1.4in,height=1.3in,clip,keepaspectratio]{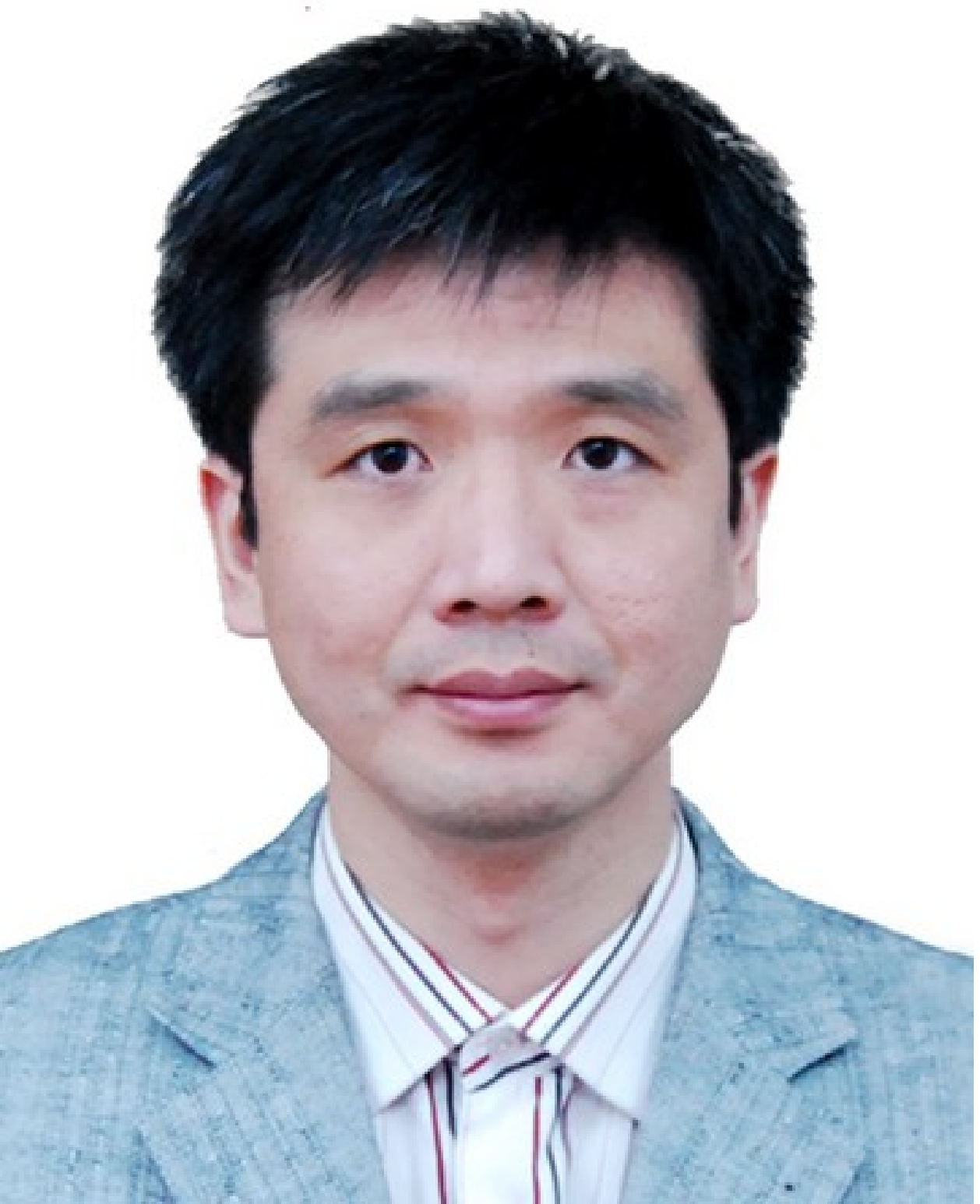}}]{Tao Han}
 (M'13) received the Ph.D. degree in communication and information engineering from Huazhong University of Science and Technology (HUST), Wuhan, China in December, 2001. He is currently an Associate Professor with the School of Electronic Information and Communications, HUST. From August, 2010 to August, 2011, he was a Visiting Scholar with University of Florida, Gainesville, FL, USA, as a Courtesy Associate Professor. His research interests include wireless communications, multimedia communications, and computer networks. Dr. Han is currently serving as an Area Editor for the EAI Endorsed Transactions on Cognitive Communications.
\end{IEEEbiography}


\begin{thebibliography}{1}

\bibitem{Thompson14}
J. Thompson, X. Ge, H. C. Wu, et. al., ``5G wireless communication systems: Prospects and challenges," {\em IEEE Commun. Mag.}, vol. 52, No. 2, pp. 62--64, Feb. 2014.

\bibitem{Boccardi14}
X. Ge, H. Cheng, M. Guizani, T. Han,  ``5G wireless backhaul networks: challenges and research advances," {\em IEEE Netw.}, vol. 28, No. 6, pp. 6--11, Nov. 2014.

\bibitem{Swindlehurst14}
X. Ge, S. Tu, G. Mao, et. al., ``5G Ultra-Dense Cellular Networks," {\em IEEE Wireless Commun.}, vol. 23, No. 1, Feb. 2016. [online]. Available: http://arxiv.org/pdf/1512.03143v1.pdf.

\bibitem{cxwang14}
C.-X. Wang, F. Haider, X. Gao, et. al., ``Cellular architecture and key technologies for 5G wireless communication networks," {\em IEEE Commun. Mag.}, vol. 52, no. 2, pp. 122--130, Feb. 2014.


\bibitem{Chen15}
M. Chen, Y. Zhang, Y. Li, S. Mao, V. Leung, ``EMC: emotion-aware mobile cloud computing in 5G", {\em IEEE Network}, Vol. 29, No. 2, pp. 32--38, Mar. 2015.

\bibitem{Chen15_2}
M. Chen, Y. Zhang, L. Hu, T. Taleb, Z. Sheng, ``Cloud-based wireless network: virtualized, reconfigurable, smart wireless network to enable 5G technologies ", {\em ACM/Springer Mobile Networks and Applications}, Vol. 20, No. 6, pp. 704--712, Dec. 2015.

\bibitem{Marzetta10}
T. L. Marzetta, ``Noncooperative cellular wireless with unlimited numbers of base station antennas," {\em IEEE Trans. Wireless Commun.}, vol. 9, no. 11, pp. 3590--3600, Nov. 2010.

\bibitem{Vu07}
M. Vu, A. Paulraj, ``MIMO wireless linear precoding," {\em IEEE Signal Proc. Mag.}, vol. 24, no. 5, pp. 86--105, Sept. 2007.

\bibitem{Zhang05}
X. Zhang, A. F. Molisch, S. Y. Kung, ``Variable-phase-shift-based RF-baseband codesign for MIMO antenna selection," {\em IEEE Trans. Signal Proc.}, vol. 53, No. 11, pp. 4091--4103, Nov. 2005.

\bibitem{Emil15}
E. Bj{\"o}rnson, L. Sanguinetti, J. Hoydis, ``Optimal design of energy-efficient multi-user MIMO systems: Is massive MIMO the answer?" {\em IEEE Trans. Wireless Commun.}, vol.
PP, no. 99, pp. 1536--1276, Feb. 2015.

\bibitem{El14}
O. E. Ayach, S. Rajagopal, S. Abu-Surra, ``Spatially sparse precoding in millimeter wave MIMO systems," {\em
IEEE Trans. Wireless Commun.}, vol. 13, no. 3, pp. 1499--1513, Jan. 2014.

\bibitem{Bogale14}
T. E. Bogale, L. B. Le, ``Beamforming for multiuser massive MIMO systems: Digital versus hybrid analog-digital," in {\em Proc. IEEE GLOBECOM 2014}, Dec. 2014, pp. 10--12.

\bibitem{Alkhateeb14}
A. Alkhateeb, O. E. Ayach, G. Leus, R. Heath, ``Channel estimation and hybrid precoding for millimeter wave cellular systems," {\em IEEE J. Sel. Topics in Signal Proc.}, vol. 4, no. 5, pp. 831--846, July 2014.

\bibitem{Ahmed14}
A. Alkhateeb, O. E. Ayach, G. Leus, R. W. Heath, ``Limited feedback hybrid precoding for multi-user millimeter wave systems," 2014 [online]. Available: http://arxiv.org/pdf/1409.5162v2.pdf.

\bibitem{Liu14}
A. Liu, V. Lau, ``Phase only RF precoding for massive MIMO systems with limited RF chains," {\em IEEE Trans. Signal Proc.}, vol. 62, no. 17, pp. 4505--4515, July 2014.

\bibitem{Liang14}
L. Liang, W. Xu, X. Dong, ``Low-complexity hybrid precoding in massive multiuser MIMO systems," {\em IEEE Wireless Commun. Letters}, vol. 3, no. 6, pp. 653--656, Oct. 2014.

\bibitem{Tadilo14}
T. E. Bogale, L. Bao Le, A. Haghighat, ``Hybrid analog-digital beamforming: How many RF chains and phase shifters do we need?" 2014 [online]. Available: http://arxiv.org/pdf/1410.2609v1.pdf.

\bibitem{Xiang13}
X. Ge, X. Huang, Y. Wang, et. al., ``Energy efficiency optimization for MIMO-OFDM mobile multimedia communication systems with QoS constraints," {\em IEEE Trans. Vehicular Tech.}, vol. 63, no. 5, pp. 2127--2138, June 2014.

\bibitem{Belmega11}
    E. V. Belmega, S. Lasaulce, ``Energy-efficient precoding for multiple-antenna terminals," {\em IEEE Trans. Signal Proc.}, vol. 59, no. 1, pp. 329--340, Jan. 2011.

\bibitem{Jiang13}
     C. Jiang, L. Cimini, ``Energy-efficient transmission for MIMO interference channels," {\em IEEE Trans. Wireless Commun.}, vol. 12, no. 6, pp. 2988--2999, May 2013.

\bibitem{Xu13}
    J. Xu, L. Qiu, ``Energy efficiency optimization for MIMO broadcast channels," {\em IEEE Trans. Wireless Commun.}, vol. 12, no. 2, pp. 690--701, Feb. 2013.


\bibitem{Ngo13}
    H. Q. Ngo, E. G. Larsson, T. L. Marzetta, ``Energy and spectral efficiency of very large multiuser MIMO systems," \emph{IEEE Trans. Commun.}, vol. 61, no. 4, pp. 1436--1449, April 2013.

\bibitem{Ha13}
D. Ha, and K. Lee, J. Kang, ``Energy efficiency analysis with circuit power consumption in massive MIMO systems," in {\em IEEE PIMRC 2013}, Sept. 2013, pp. 938--942.

\bibitem{Mohammed14}
    S. Mohammed, ``Impact of transceiver power consumption on the energy efficiency of zero-forcing detector in massive MIMO systems," {\em IEEE Trans. Wireless Commun.}, vol. 62, no. 11, pp. 3874--3890, Oct. 2014.

\bibitem{Yang13}
    H. Yang, T. L. Marzetta, ``Total energy efficiency of cellular large scale antenna system multiple access mobile networks," in {\em Proc. IEEE Online Conference on Green Communications (GreenCom)}, pp. 27--32, Oct. 2013.

\bibitem{Ng12}
    D. W. K. Ng, E. S. Lo, R. Schober, ``Energy-efficient resource allocation in OFDMA systems with large numbers of base station antennas," \emph{IEEE Trans. Wireless Commun.}, vol. 11, no. 9, pp. 3292--3304, Sep. 2012.

\bibitem{Xu02}
    H. Xu, V. Kukshya, T. Rappaport, ``Spatial and temporal characteristics of 60-GHz indoor channels," {\em IEEE J. Sel. Areas Commun.}, vol. 20 no. 3, pp. 620--630, Apr. 2002.

\bibitem{Raghavan11}
V. Raghavan, A. M. Sayeed, ``Sublinear capacity scaling laws for sparse MIMO channels," {\em
IEEE Trans. Information Theory}, vol. 57, no 1, pp. 345--364, Jan. 2011.

\bibitem{Balanis12}
C. A. Balanis, {\em Antenna Theory: Analysis and Design}. John Wiley and Sons, 2012.

\bibitem{Bogale14CISS}
T. E. Bogale, L. B. Le, ``Pilot optimization and channel estimation for multiuser massive MIMO systems," in {\em Proc. IEEE Conference on
Information Sciences and Systems}, Mar. 2014, pp. 1--6.

\bibitem{Bajwa10}
W. U. Bajwa, J. Haupt, A. M. Sayeed, and R. Nowak, ``Compressed channel sensing: a new approach to estimating sparse multipath channels," {\em Proc. IEEE}, vol. 98, no. 6, pp. 1058--1076, 2010.

\bibitem{Boyd04}
 S. Boyd, L. Vandenberghe, {\em Convex optimization}, Cambridge university press, 2004.



\bibitem{Jiang11}
C. Jiang, L. J. Cimini, ``Downlink energy-efficient multiuser beamforming with individual SINR constraints", in {\em Proc. MILCOM 2011}.

\bibitem{Lee12}
J. M. Lee, {\em Introduction to smooth manifolds}, Springer Science \& Business Media, 2012.

\bibitem{Golub96}
G. Golub and C. Van Loan, {\em Matrix Computations, 3rd ed}, The John Hopkins Univ. Press, 1996.

\bibitem{He14}
S. He, Y. Huang, L. Yang, et. al., ``Coordinated multicell multiuser precoding for maximizing weighted sum energy efficiency", {\em
IEEE Trans. Signal Proc.}, vol. 62, no. 3, pp 741--751, Feb. 2014.


\bibitem{Sulyman14}
A. I. Sulyman, A. T. Nassar, M. K. Samimi, et. al., ``Radio propagation path loss models for 5G cellular networks in the 28 GHZ and 38 GHZ millimeter-wave bands", {\em Commun. Mag.}, vol. 52, no. 9, pp. 78--86, Sept. 2014.

\bibitem{Akdeniz14}
M. R. Akdeniz, Y. Liu, Samimi, M. K. Samimi, et al.et. al., ``Millimeter wave channel modeling and cellular capacity evaluation," {\em IEEE J. Sel. Areas Commun.}, pp. 1164--1179, June 2014.


\bibitem{Pei12}
L. Xiang, X. Ge, C.-X. Wang, F. Y. Li, and F. Reichert, ``Energy efficiency evaluation of cellular networks based on spatial distributions of traffic load and power consumption," {\em IEEE Trans. Wireless Commun.}, vol. 12, no. 3, pp. 961--973, Mar. 2013.

\bibitem{Yong07}
    S. K. Yong, and C. C. Chong, ``An overview of multigigabit wireless through millimeter wave technology: potentials and technical challenges", {\em EURASIP Journal on Wireless Communications and Networking}, vol. 2007, no. 1, pp. 50--50, Jan. 2007.


\bibitem{Kalivas95}
    G. Kalivas, M. El-Tanany and S. Mahmoud, ``Millimeter-wave channel measurements with space diversity for indoor wireless communications", {\em IEEE Trans. Vehicular Tech.}, vol. 44, no. 3, pp. 494--505, Aug. 1995.

\bibitem{Byers88}
 R. Byers, ``A bisection method for measuring the distance of a stable matrix to the unstable matrices," {\em SIAM Journal on Scientific and Statistical Computing}, vol. 9, no. 5, pp. 875--881, Feb. 1988.

\bibitem{Varma13}
  V. S. Varma, M. Debbah, S. E. Elayoubi, ``An energy-efficient framework for the analysis of MIMO slow fading channels," \emph{IEEE Trans. Signal Proc.}, vol. 61, no. 10, pp. 2647--2659, March 2013.


\bibitem{Tombaz11}
    S. Tombaz, A. Vastberg, J. Zander, ``Energy- and cost-efficient ultra-high-capacity wireless access," \emph{IEEE Wireless Commun. Mag.}, vol. 18, no. 5, pp. 18--24, October 2011.

\bibitem{Cui04}
    S. Cui, A. Goldsmith, A. Bahai, ``Energy-efficiency of MIMO and cooperative MIMO techniques in sensor networks," \emph{IEEE J. Sel. Areas Commun.}, vol. 22, no. 6, pp. 1089--1098, August 2004.


\end{thebibliography}
\end{document}